\documentclass[useAMS,usenatbib]{mn2e}

\usepackage{lscape}
\usepackage{tabularx}
\usepackage{xtab}
\usepackage{xspace}
\usepackage{graphicx}
\usepackage{placeins}
\usepackage{amsfonts,amsmath,amssymb}
\usepackage{url}
\usepackage[usenames,dvipsnames]{color}
\usepackage{times}

\usepackage{enumerate}
\include{JournalAbbr}    
\newcommand{\msun}{~\mathrm{M}_{\odot}}

\title[Zooming in on major mergers]{Zooming in on major mergers: dense, starbursting gas in cosmological simulations}

\author[Sparre \& Springel]{\parbox[t]{\textwidth}{
		Martin Sparre$^{1,2}$\thanks{Sapere Aude Fellow, E-mail:sparre@dark-cosmology.dk} and
		Volker Springel$^{1,3}$
		\vspace*{6pt}} \\
$^1$Heidelberger Institut f{\"u}r Theoretische Studien, Schloss-Wolfsbrunnenweg 35, 69118 Heidelberg, Germany\\
$^2$Dark Cosmology Centre, Niels Bohr Institute, University of Copenhagen, Juliane Maries Vej 30, 2100 Copenhagen, Denmark\\
$^3$Zentrum f\"ur Astronomie der Universit\"at Heidelberg, Astronomisches Recheninstitut, M\"onchhofstrasse 12-14, 69120 Heidelberg, Germany\\
}

\begin{document}

\date{\today}

\pagerange{\pageref{firstpage}--\pageref{lastpage}} \pubyear{2016}
\maketitle

\label{firstpage}

\begin{abstract} We introduce the `Illustris zoom simulation project', which allows the study of selected galaxies forming in the $\Lambda$CDM cosmology with a 40 times better mass resolution than in the parent large-scale hydrodynamical Illustris simulation. We here focus on the starburst properties of the gas in four cosmological simulations of major mergers. The galaxies in our high-resolution zoom runs exhibit a bursty mode of star formation with gas consumption timescales 10 times shorter than for the normal star formation mode. The strong bursts are only present in the simulations with the highest resolution, hinting that a too low resolution is the reason why the original Illustris simulation showed a dearth of starburst galaxies. Very pronounced bursts of star formation occur in two out of four major mergers we study. The high star formation rates, the short gas consumption timescales and the morphology of these systems strongly resemble observed nuclear starbursts. This is the first time that a sample of major mergers is studied through self-consistent cosmological hydrodynamical simulations instead of using isolated galaxy models setup on a collision course. We also study the orbits of the colliding galaxies and find that the starbursting gas preferentially appears in head-on mergers with very high collision velocities. Encounters with large impact parameters do typically not lead to the formation of starbursting gas.  \end{abstract}


\section{Introduction}

Galaxies appear to form stars in at least two modes; a quiescent and a starburst mode. An observational relation revealing these modes is the Kennicutt--Schmidt relation between the gas surface density ($\Sigma_\text{gas}$) and the star formation rate per surface area ($\Sigma_\text{SFR}$) within a galaxy \citep{1998ApJ...498..541K}.  Quiescently star-forming disc galaxies typically follow a power law relation, $\Sigma_\text{SFR} \propto \Sigma_\text{gas}^{1.3}$ \citep{2012ApJ...745...69K}, whereas starburst galaxies follow a relation with a similar power index, but a $\simeq 15$ times higher normalisation, implying that gas is converted much faster into stars in these galaxies.  A key element for characterizing these modes is hence the timescale with which star formation consumes the gas within a galaxy.

In local disk galaxies, a linear relation exists between the star formation surface density, $\Sigma_\text{SFR}$, and the surface density of molecular hydrogen, $\Sigma_{\text{H}_2}$ \citep{2011ApJ...730L..13B}. A plausible interpretation of this linearity is that star formation is physically tied to the presence of molecular gas, rather than to the total gas surface density. As for the $\Sigma_\text{SFR} - \Sigma_\text{gas}$ relation, the $\Sigma_\text{SFR} - \Sigma_{\text{H}_2}$ also breaks down for starbursting galaxies such as ultra-luminous infrared galaxies \citep[ULIRGS,][]{1991ApJ...370..158S}, where molecular gas is converted into stars more rapidly than for quiescently star-forming galaxies.  While the observational picture thus clearly shows that a bursty mode of star formation exists with roughly 15 times shorter gas depletion timescales than for the normal star formation mode, it is less clear what conditions are required to trigger this accelerated star formation.

Observations also show a redshift-dependent relation between the SFR and stellar mass of galaxies \citep{2004MNRAS.351.1151B,2007ApJ...660L..43N,2014ApJS..214...15S}. Galaxies being significantly more star-forming than predicted by this relation are also often referred to as starbursts \citep{2011ApJ...739L..40R}, even though such a selection criterion does not select the exact same population of galaxies as a selection by gas depletion timescales \citep{2009ApJ...698.1437K}. The term \emph{starburst} is thus ambiguous, with some definitions relying on gas consumption timescales (of the ISM gas, or only the dense star-forming gas), while other definitions rely on the SFR$/M_*$-value of the galaxy, or alternatively on the deviation of a galaxy from the mean SFR--$M_*$ relation. Throughout this paper we define the term starburst based on the gas consumption timescale derived from the SFR and the gas mass of a galaxy.

\begin{table*}
\centering
\begin{tabular}{c|cccc|c|cccc} 
\hline\hline 
Sim. Name  &$\log \frac{M_{*,1}}{\msun}$ & $\frac{\log M_{*,2}}{\msun}$ & $M_1/M_2 $& $\log\frac{ M_*(z=0)}{\msun}$ & $\log \frac{M_{200}(z=0)}{\msun}$&$\epsilon_b/$kpc &$\epsilon_\text{dm}/$kpc &$m_b/\msun$&$m_\text{dm}/\msun$\\
\hline
\hline
1330-1 & 9.87 & 9.69 & 1.51 & 10.82 & 12.19 &  0.64 & 0.64 & $4.42\times 10^6$ & $4.42\times 10^6$\\
1330-2 & 9.99 & 9.88 & 1.27 & 10.96 & 12.19 &  0.32 & 0.32 & $5.53\times 10^5$ & $5.53\times 10^5$\\
1330-3 & 10.14 & 9.99 & 1.41 & 11.04 & 12.19 & 0.21 & 0.21 & $1.64\times 10^5$ & $1.64\times 10^5$\\
\hline
1349-1 & 9.70 & 9.68 & 1.05 & 10.73 & 12.17 &  0.64 & 0.64 & $4.42\times 10^6$ & $4.42\times 10^6$\\
1349-2 & 9.93 & 9.82 & 1.30 & 10.83 & 12.15 &  0.32 & 0.32 & $5.53\times 10^5$ & $5.53\times 10^5$\\
1349-3 & 10.00 & 10.00 & 1.00 & 10.92 & 12.15 & 0.21 & 0.21 & $1.64\times 10^5$ & $1.64\times 10^5$ \\
\hline
1526-1 & 9.78 & 9.65 & 1.34 & 10.61 & 12.26 &  0.64 & 0.64  & $4.42\times 10^6$ & $4.42\times 10^6$\\
1526-2 & 9.95 & 9.84 & 1.29 & 10.68 & 12.27 &  0.32 & 0.32 & $5.53\times 10^5$ & $5.53\times 10^5$\\
1526-3 & 10.10 & 10.03 & 1.18 & 10.75 & 12.25 & 0.21 & 0.21  & $1.64\times 10^5$ & $1.64\times 10^5$\\
\hline
1605-1 & 10.02 & 9.88 & 1.38 & 10.65 & 12.00 &  0.64 & 0.64 & $4.42\times 10^6$ & $4.42\times 10^6$\\
1605-2 & 10.19 & 10.08 & 1.30 & 10.81 & 12.01 &  0.32 & 0.32 & $5.53\times 10^5$ & $5.53\times 10^5$ \\
1605-3 & 10.27 & 10.21 & 1.16 & 10.89 & 12.00 & 0.21 & 0.21  & $1.64\times 10^5$ & $1.64\times 10^5$\\
\hline\hline
\end{tabular}
  \caption{An overview of the merger simulations presented in this paper. Four different mergers from the Illustris simulation are simulated at three different resolutions, so a total of 12 simulations have been run. The simulations are named as AAAA-B, where AAAA is the $z=0$ friends-of-friends group number in Illustris \citep[see the public data release of Illustris,][]{2015A&C....13...12N}, and B is the `zoom factor'. $M_{*,1}$ ($M_{*,2}$) is the stellar mass at $z=0.93$ of the largest (smallest) galaxy participating in the merger. $M_{*,1}/M_{*,2}$ is the mass ratio of the two galaxies. $M_*(z=0)$ is the $z=0$ stellar mass of the merger remnant, and $M_{200}(z=0)$ is the mass enclosed in a sphere with a density 200 times the critical density of the universe. $\epsilon_b$ ($\epsilon_\text{dm}$) is the maximum physical softening length of the baryon (dark matter) particles. At $z\leq 1$, the softening is fixed to these physical values, and at $z>1$ the softening is fixed to the co-moving values at $z=1$. $m_b$ and $m_\text{dm}$ are the masses for the baryonic cells and dark matter particles, respectively.}
\label{SimOverview}
\end{table*}

Hydrodynamical simulations of isolated major mergers can account for the high star formation rates seen in starbursting ULIRGs \citep{1994ApJ...431L...9M,1996ApJ...464..641M,2005MNRAS.361..776S,2014MNRAS.442.1992H}. Radiative transfer analysis of major merger simulations has also reproduced their high submillimeter fluxes \citep{2010MNRAS.401.1613N}. Most of the literature studying major mergers has however employed idealized setups, where the initial conditions were equilibrium disc galaxies and the galaxy orbits have usually been collided on zero-energy orbits. Additionally, ad-hoc choices for the relative orientation of the galaxy discs and for the amount of orbital angular momentum have to be made. An exception to this idealised approach is the work by \citet{2009MNRAS.398..312G}, who analyzed the morphological evolution of a major merger at $z<1$, finding that the $z=0$ merger remnant had a significant disc surrounding a central bulge.

The hydrodynamical simulation Illustris is a large-scale cosmological simulation in which many major mergers occur \citep{2015MNRAS.449...49R}. The simulation has shown interesting successes in describing the evolution of star-forming galaxies \citep{2014MNRAS.445..175G,2014MNRAS.444.1518V}, and in the formation of elliptical and spiral galaxies \citep{2014Natur.509..177V, 2015MNRAS.454.1886S, 2015MNRAS.447.2753T}. A short-coming of the simulation is, however, that it did not produce enough starburst galaxies with a very high star formation rate compared to the rest of the galaxy population \citep{2015MNRAS.447.3548S}. A potential reason for this deficit of galaxies with large SFRs is that the $\simeq 700$ pc resolution of Illustris is not fine enough to resolve the central starbursting regions within galaxies sufficiently well.

The aim of this paper is to study the starbursting gas in high-resolution zoom-in simulations of major mergers identified in the Illustris simulation. These simulations are cosmological, self-consistent simulations, unlike the majority of the major merger calculations in the literature, and hence remove the reliance on idealised initial conditions and prescribed orbital parameters. We select individual major merger remnants from Illustris and repeat the simulation with spatially variable resolution, allowing us to zoom in on the galaxies of interest with higher resolution. With this approach it is possible to study selected galaxies at a 40 times better mass-resolution than in Illustris. Our analysis focusses on the conditions under which starbursting gas with a short consumption timescale appears in galaxy mergers. We will also discuss whether the reason for the lack of starburst galaxies in the original Illustris run is caused by resolution limitations for resolving the starbursting gas \citep[as suggested in][]{2015MNRAS.447.3548S}.

This paper is structured as follows. In Section~\ref{methods}, we describe our initial conditions and the physical galaxy formation model, and in Section~\ref{GasSection} we discuss the gas and star formation properties of the galaxies before, during and after the starburst associated with the major mergers. In Section~\ref{OrbitSection}, we assess the orbit of the mergers, and see how it affects the starburst properties of the mergers. Finally, in Section~\ref{Discussion} we discuss our results and in Section~\ref{Conclusion} we summarize our conclusions.

\section{Methods} \label{methods}

\subsection{Galaxy formation model}\label{PhysModel}

We use the same physical model as for the Auriga simulation project \citep[first presented in][]{2015arXiv151202219G}. This model is closely based on \citet{2014MNRAS.437.1750M} and \citet{2013MNRAS.436.3031V}. The hydrodynamical calculations are carried out with the moving mesh code AREPO (\citealt{2010MNRAS.401..791S}, see also \citealt{2016MNRAS.455.1134P}), and star formation and winds are described with the \citet{2003MNRAS.339..289S} model that treats the interstellar medium (ISM) with a subgrid approach, where the star-forming clouds are not resolved, but instead each star-forming gas cell has an assigned fraction of cold gas and a hot ambient phase based on a simple analytic prescription.
Stars form out of the cold phase, and the energy released by supernova type II is assumed to reheat some of the cold gas and transfer it to the hot gas. Together with the evaporation of cold gas trough thermal conduction, this creates a pressurized star forming medium.

Star formation occurs in gas cells with a density above $\rho_\text{th}=0.13$ cm$^{-3}$. The star formation timescale of a gas cell is set to $t_* = 2.27 \text{ Gyr} \times (\rho/\rho_\text{th})^{-0.5}$, so it is proportional to the local dynamical time. Stellar population particles are used to represent single age stellar populations with an initial mass function from \citet{2003PASP..115..763C}. Mass loss and chemical enrichment of these stellar population particles are tracked continuously. When mass is returned to the ISM, the mass in various elements (H, He, C, N, O, Ne, Mg, Si and Fe) is distributed to the 64 nearest gas cells.

Kinetic feedback from SN type-II in the form of winds is created by probabilistically converting star-forming gas cells into wind particles. The winds have an initial speed that scales with the local dark matter velocity dispersion, which gives a remarkably better fit to the SFR--$M_*$ than, e.g., a constant wind velocity \citep{2006MNRAS.373.1265O, 2011MNRAS.415...11D, 2013MNRAS.428.2966P}. The winds are briefly decoupled from the hydrodynamical interactions after their release, until they reach a gas density slightly below the star formation threshold. This decoupling and recoupling procedure ensures that the winds escape the star-forming regions without disrupting them, and that the winds actually attain the prescribed mass-loading factor. When a wind particle recouples, its momentum, thermal energy and metal content is dumped into the nearest gas cell, so that the winds effectively constitute a non-local feedback that occurs at the transition layer between star-forming and non-starforming gas.

Black holes are seeded when a halo reaches a mass of $7.1\times10^{10} \msun$, and the adopted seed mass is $1.4\times10^5 \msun$. The black holes grow either by Bondi-Hoyle or Eddington-limited accretion (the smaller of the two rates applies). The feedback \emph{mode} of the black hole is determined by the accretion rate \citep{2007MNRAS.380..877S} relative to the Eddington rate. For high accretion rates the black hole is assumed to be in the \emph{quasar mode}, where thermal feedback heats the gas in the vicinity of the black hole, while in the low-accretion mode, feedback acts mechanically in a \emph{radio mode}, which is here implemented as a distributed heating of the halo gas.

Furthermore, we use a uniform UV background \citep{2009ApJ...703.1416F}, and gas cooling rates (see \citealt{2013MNRAS.436.3031V} for details) with self-shielding corrections from \citet{2013MNRAS.430.2427R}. More details of the physical models will be presented in the main Auriga project paper (Grand et al. in prep.).

\subsection{Four major mergers from the Illustris simulation} \label{MergerSelection}

Initial conditions are created by selecting galaxies from the Illustris simulation \citep{2014Natur.509..177V, 2014MNRAS.444.1518V,2014MNRAS.445..175G} at $z=0$ that experienced a major merger between $z=1$ and $z=0.5$. By selecting galaxy mergers in this redshift interval we ensure that the mergers have time to finish before $z=0$, so we can study, e.g., how the relaxed stellar system at $z=0$ is affected by the merger. We base our selection on the merger trees from Illustris.

We have created zoom initial conditions (ICs) for 4 major mergers using a modified version of the N-GenIC code \citep{2015ascl.soft02003S} that had also been used to create the ICs of the Illustris simulation. The code coarsens the resolution of the initial particle load progressively with distance from the galaxy of interest, while increasing the resolution inside the Lagrangian region of the galaxy and its immediate surroundings. In the resulting setup, we distinguish three types of dark matter particles (DM): the high resolution particles, the standard resolution particles (with resolution very close to the original Illustris resolution) and the low resolution particles (further away from the target galaxy). In the initial particle load, the standard resolution particles form a shell around a roughly spherical high-resolution region, with the low resolution particles filling the rest of the simulation box. For each particle type, the displacement field is sampled in Fourier space up to the Nyquist frequency of the corresponding particle load. All waves that were present already in Illustris are kept so that the same object forms again, but augmented with additional small-scale power in the initial conditions. 

The standard resolution particles in our setup have a mass of $M_\text{tot,dm} / 2048^3$ (here $M_\text{tot,dm}$ is the total DM mass in the box), and the high resolution particles have a mass which is lowered by a `zoom factor' cubed.  The low resolution particles have a variable mass that depends on the distance to the galaxy of interest. Baryons are added to the simulation by putting the cosmologically expected fraction of the mass of each dark matter particle into baryonic gas cells. The corresponding mesh generating points and dark matter particles are displaced relative to each other such that the center-of-mass of each dark matter and cell pair is unchanged, and the mean distance between dark matter and mesh generating points is maximized. This is done to avoid artificial pairing effects if a small gravitational softening length is adopted.

The finest dark matter mass resolution level is set by the `zoom factor' according to the formula,
\begin{align*}
m_\text{dm}=\left(\frac{1820}{2048\times \text{`zoom factor'}}\right)^3 \times  6.299 \times 10^6\msun.
\end{align*}
The factor of 1820/2048 is the ratio between the number of dark matter particles per dimension in Illustris, and our standard resolution. For each galaxy we ran simulations with a zoom factor of 1, 2 and 3, which corresponds to mass resolutions 1.4, 11.4 and 38.5 times finer than in the Illustris simulation. The gravitational softening was chosen to be $\epsilon \simeq \frac{L}{40 N\times \text{`zoom factor'}}$, where $L=106.25\,{\rm Mpc}$ is the width of the cosmological box, and $N$ is the number of particles per dimension for the standard resolution particles.  Our dark matter mass resolution for the runs with a `zoom factor' equal to 3 is therefore 1.7 times coarser than the ERIS simulation \citep{2011ApJ...742...76G} and $\gtrsim 7$ times finer than in the Milky Way simulations presented by \citet{2014MNRAS.437.1750M}. In our simulation, the softenings are fixed to this physical value for $z\leq1$. At $z>1$ the comoving softening is fixed to the (comoving) value at $z=1$.

\begin{figure*}
\centering
\includegraphics[width = 0.9 \textwidth]{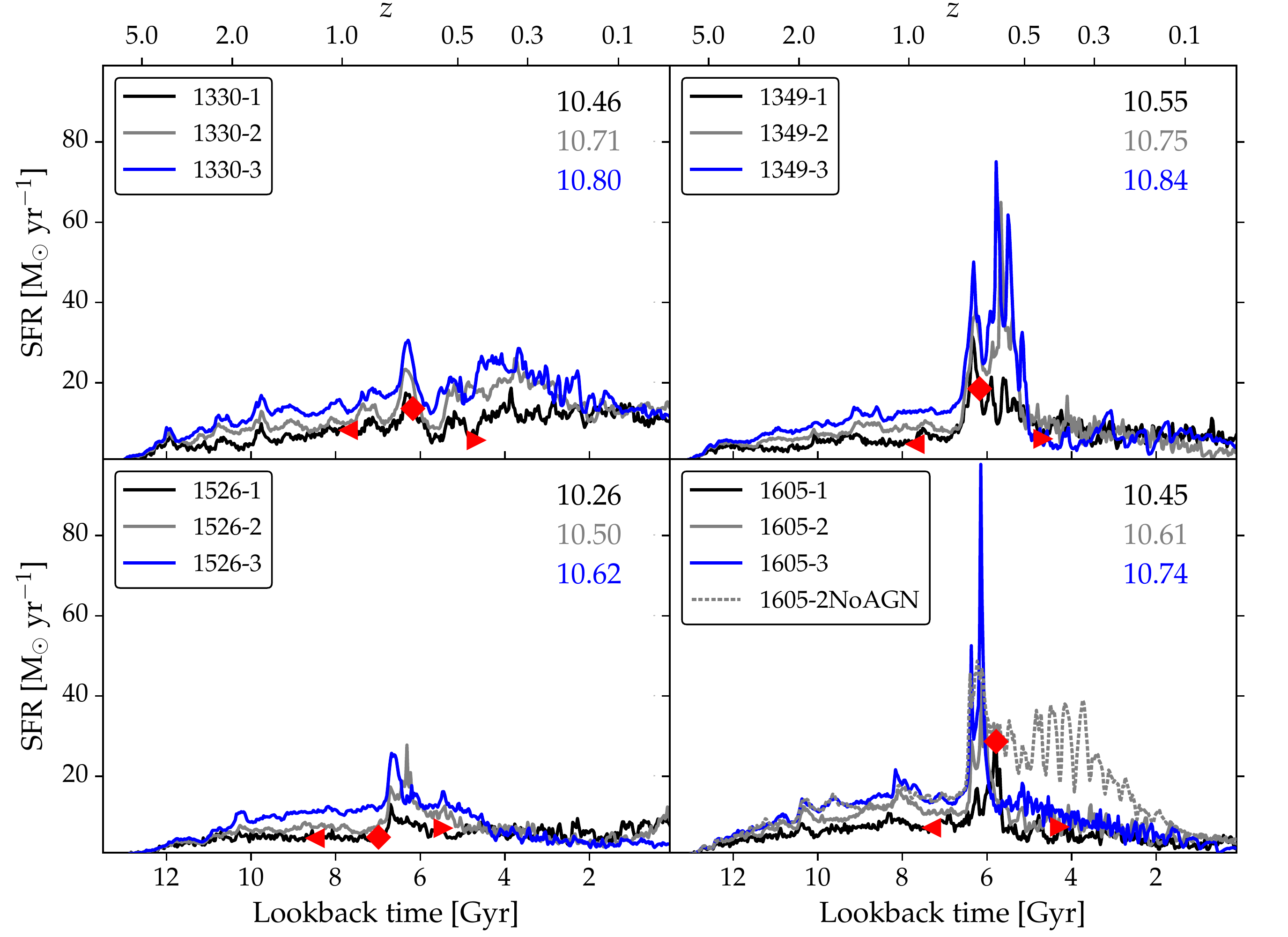}
\caption{The star formation histories of all the major mergers calculated based on the stars present in each galaxy at $z=0$. Each panel shows a galaxy simulated at either zoom factor 1 (\emph{black}), 2 (\emph{grey}) or 3 (\emph{blue}), with resolution level 3 being the finest resolution and resolution level 1 being the coarsest. The merging time, which is marked with a $\Diamond$-symbol, is defined to be at the peak of the specific black hole accretion rate, $\dot{M}_\text{BH}/M_\text{BH}$, and the start/end of the merger is defined to be 1.5 Gyr before/after the merger (see $\triangleleft/\triangleright$ symbols). In the \emph{upper right corner} the logarithm of the stellar mass formed between the start and end of the merger is tabulated (`zoom factor' 1, 2 and 3 correspond to the \emph{top}, \emph{middle} and \emph{bottom row}, respectively). In all cases, a larger amount of stellar mass is formed during the merger when the resolution is increased. The \emph{grey-dashed line} in the \emph{lower right panel} shows a simulation of the 1605-2 galaxy without black holes and AGN feedback.}
\label{AllSFR}
\end{figure*}

The main properties of the four major merger simulations are summarized in Table~\ref{SimOverview}. We list the mass of the most-massive galaxy ($M_1$), and the second most-massive galaxy ($M_2$) participating in the major merger at $z=0.93$, which is before the collisions occur. The mass ratios of all the major mergers are between 1:1.05 and 1:1.51. Before the mergers, the progenitors have masses in the range $10^{9.65}-10^{10.27}\msun$, and at $z=0$ the merger remnants have reached masses of $10^{10.61}-10^{11.04}\msun$. At $z=0$, the galaxies reside in host halos with masses of $M_{200}=10^{12.00}-10^{12.27}\msun$, where $M_{200}$ is the mass within a sphere centered on the halo containing a mean density of 200 times the critical density of the universe. All our galaxies therefore have masses close to the Milky Way at the present epoch, both in terms of $M_*$ and $M_{200}$ \citep[according to recent constraints the Milky Way has a stellar mass of $M_*=6.43\pm 0.63 \times 10^{10}\msun$ and a virial mass of $1.26\pm 0.24 \times 10^{12}\msun$; ][]{2011MNRAS.414.2446M}.

\subsection{Tracking the main galaxy}

Each simulation contains two massive galaxies before the merger, while after the merger event one massive galaxy is left. A significant part of our analysis in this paper relies on merger trees where we track the galaxy in each simulation with the largest stellar mass at $z=0.93$ immediately before the merger. To this end we store the ID of each star in the galaxy, and at any other redshift, the main descendant or progenitor of the galaxy is identified with the object that has the largest number of stars in common with the $z=0.93$ galaxy.

\section{Star formation histories and gas consumption timescales} \label{GasSection}

\subsection{Star formation rate enhancements during starbursts}\label{SFHsubsection}

In Figure~\ref{AllSFR}, we plot the star formation histories of our simulated galaxies based on the formation time and initial mass of the star particles found at $z=0$ in the remnant galaxies. The star formation rate (SFR) has been calculated in bins of width 28 Myr. Each panel shows a galaxy simulated at different resolution, with mass resolutions 1.4, 11.4 and 38.5 times finer than in the Illustris simulation. Common for all simulations is that one or several peaks in the SFR history are present at lookback times between 5 and 8 Gyr, which is where the merger happens according to the selection criteria in Section~\ref{MergerSelection}. For each star formation history, the `merging time', $t_\text{merge}$, defined to be the time with the largest black hole accretion rate in the redshift-interval $0.5<z<1.0$ is marked. Furthermore, we define the start and end time ($t_\text{start}$ and $t_\text{end}$) of the merger phase as 1.5 Gyr before and after this merging time, respectively. Compared to the SFR-peak, which appears during the merger (i.e.~with $t_\text{start}<t<t_\text{end}$), the galaxies evolve quiescently before and after the merger.

An interesting feature is that the high-resolution runs have larger SFR-peaks than the low resolution runs. And it is not only the merger-induced SFR-peak that is larger for the high-resolution runs, but also the integrated SFR during the merger, which is $\gtrsim 2$ times larger for the highest compared to the lowest resolution run (the amount of stellar mass formed within 1.5 Gyr of the burst is noted in the upper right corner of each panel). This result -- that an increased resolution of the simulation leads to more stars formed during a merger-induced starbursts -- potentially resolves the problem with too little star formation occurring in starbursts in the Illustris simulation \citep{2015MNRAS.447.3548S}. At $z=0$, the stellar mass present in the high-resolution runs is also 0.14-0.24 dex larger than for the low resolution runs, see Table~\ref{SimOverview}. 

\begin{figure*}
\centering
\includegraphics[width = 0.9 \textwidth]{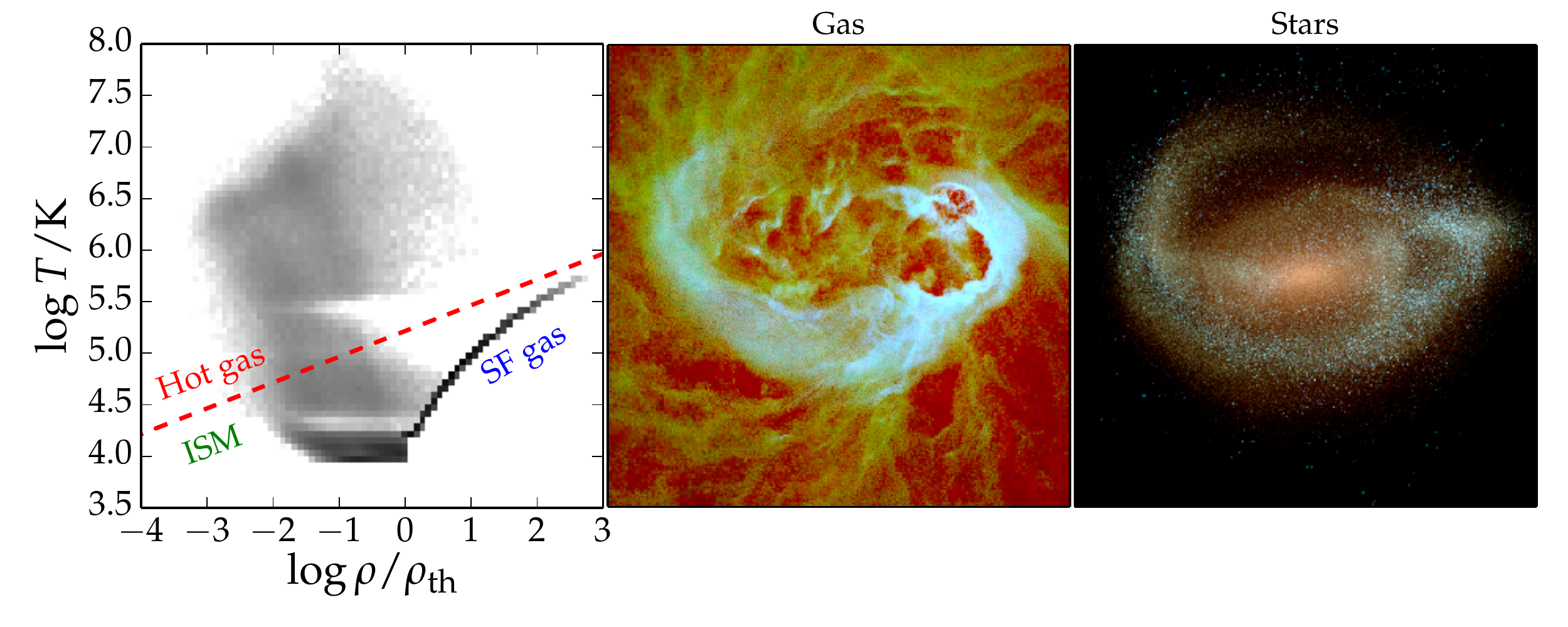}
\caption{A face-on view of the different gas phases shown for the 1330-3 galaxy at $z=0$. The \emph{left panel} shows how gas is divided into a hot and an ISM phase (above and below the \emph{dashed line}, respectively), and also shown is the star-forming part of the ISM. The \emph{central panel} shows the spatial distribution of the different gas phases. The hot gas is shown as \emph{red}, the ISM is \emph{green}, and the star-forming gas is \emph{blue} (and appears \emph{blue-white}). The ISM gas is located in a ring/disc-like structure, and the star-forming part of the ISM is in the densest parts of this disc. The hot gas surrounds the disc. The \emph{right panel} shows the distribution of stars (\emph{blue}, \emph{green} and \emph{red} represent the $U$, $B$ and $K$ color, respectively), which have a similar ring/disc like morphology as the gas. In the center of the galaxy is a gas-poor bulge with a red stellar component.}
\label{RhoT}
\end{figure*}

This difference in stellar mass is smaller than the difference in stellar mass formed during the merger epoch ($0.29-0.36$ dex, the numbers are shown in the upper right part of each panel in Figure~\ref{AllSFR}). This is because the galaxies have a more similar star formation history after the merger is finished than during the merger. The 1605-1, 1605-2 and 1605-3 simulations, for example, illustrate this in a clear way: for lookback times larger then 5 Gyr, the SFR is roughly two times larger for the best resolved simulations compared to the low resolution runs. At smaller lookback times these differences go away, and at a lookback time of 2 Gyr, the SFR is essentially the same in all the simulations. At $z=0$ there is actually a slightly larger SFR in the 1605-1 run than in the 1605-3 run, which is most likely because the latter simulation used up more of the gas available for star formation earlier in its history.

In idealized mergers, \citet{2006MNRAS.373.1013C} found a similar result, namely that increasing the resolution in major mergers also increased the peak star formation. Their galaxies showed, however, a more similar behaviour in the pre- and post-burst phases than our cosmological simulations. Various explanations are possible for these differences. First, it is possible that a self-consistent inclusion of cosmological structure formation -- as done in our simulations -- might give more star formation on smaller scales compared to the idealized simulations, where the disc initially follows a smooth mass profile. Second, it is also possible that our hydrodynamical scheme is better at resolving, e.g., subsonic turbulence than the smoothed particle hydrodynamics method used in the idealized simulations \citep{2012MNRAS.423.2558B}. Finally, the simulations of \citet{2006MNRAS.373.1013C} also adopted a different physical galaxy formation model that did not include feedback from AGN, which we speculate could change the convergence behaviour of our simulations. \citet{2014MNRAS.442.1992H} also studied idealized mergers and found that the star formation peaks are quite sensitive to changes of resolution, the numerical hydrodynamical scheme, and whether or not black hole feedback was enabled. This is consistent with our interpretation.

Especially at high redshifts in the range $1\leq z\leq 2$, the star formation rates of the galaxies are not fully converged; the high-resolution simulations typically have up to 2 times higher SFR than in the low-resolution simulations. Similar convergence issues were seen in the $25 \, h^{-1}\,{\rm Mpc}$ box simulations of \citet{2013MNRAS.436.3031V}  performed with the Illustris physics model. In~Section~\ref{SFRRho} we will elaborate on this further.

To estimate the role of AGN feedback we have performed a simulation of the 1605-2 galaxy without black hole formation and AGN feedback (\emph{grey-dashed line} in the \emph{lower right panel} of Figure~\ref{AllSFR}). The run without black holes have an up to two times larger SFR during and before the merger. This is not surprising given that the black hole is responsible for heating and removing gas in the center of galaxies. In the post-merger regime the ratio between the SFR in the run with and without black holes is larger. This is expected because a black hole grows rapidly during a merger. The figure hence reveals that the feedback associated with black holes is crucial for determining the post-merger evolution, but it plays a slightly minor role before and during the merger.

\subsection{Dividing the gas into a hot and an ISM phase}\label{ISMExplanation}

Before studying the starburst properties of major mergers we now discuss how we separate the gas in halos into an ISM phase, which contains cold gas and where star formation occurs, and a hot phase, which consists of gas heated by accretion shocks and stellar winds from feedback effects associated with star formation. Distinguishing these phases is typically based on the position of a gas cell in the density--temperature diagram. An example is given in Figure~\ref{RhoT} (\emph{left panel}), where we have constructed a two-dimensional histogram using a $80\times 80$ grid in the $\rho$--$T$ plane for the 1330-3 galaxy. The logarithmically scaled intensity of the grey color encodes the amount of mass in each cell of the histogram. The range of mass in the (grey) non-empty histogram cells covers five orders of magnitudes. The behaviour of the gas in this diagram is closely linked to the galaxy formation model. The thin one-dimensional curve starting at $\rho=\rho_\text{th}$ and extending to higher densities is the location of the star-forming gas, which follows the effective equation of state resulting from our subgrid-physics model. As described in Section~\ref{PhysModel}, star formation occurs most rapidly at high densities.

Formally, we will base our definition of the hot and the ISM gas on the \emph{the dashed line}, which takes the form
\begin{align*}
\log \left( \frac{T}{\text{K}} \right)< 6+0.25 \log \left(\frac{\rho}{10^{10}h^2\msun \text{kpc}^{-3}}\right),
\end{align*}
as suggested by \citet{2012MNRAS.427.2224T}. We will generally refer to the gas above this relation in the $\rho-T$ diagram as \emph{hot gas}, and the gas below as ISM gas (with mass $M_\text{ISM}$). The mass of the star-forming gas, which is a subset of the ISM gas, is denoted by $M_\text{SFR}$ in the remaining part of this paper.

The hot gas is mostly located at lower densities and higher temperatures than the star-forming gas. Important processes for heating the gas to such high temperatures are accretion shocks, which heat gas to the virial temperature of a halo, and heating by feedback. The part of the ISM which is not star-forming is at similar densities as the hot gas, but with a much lower temperature. The spatial distribution of the gas phases is shown in the \emph{central panel}. Here the hot gas is shown in \emph{red}, the ISM gas in \emph{green} and the star-forming gas in blue. A logarithmic scale is used for all of the colors. With our chosen color-coding the star-forming gas appear \emph{blue--white} and is the densest part of the ISM gas, making up a disc-like structure. The hot gas on the other hand dominates outside of the disc.

\begin{figure*} \centering \includegraphics[width = 0.9 \textwidth]{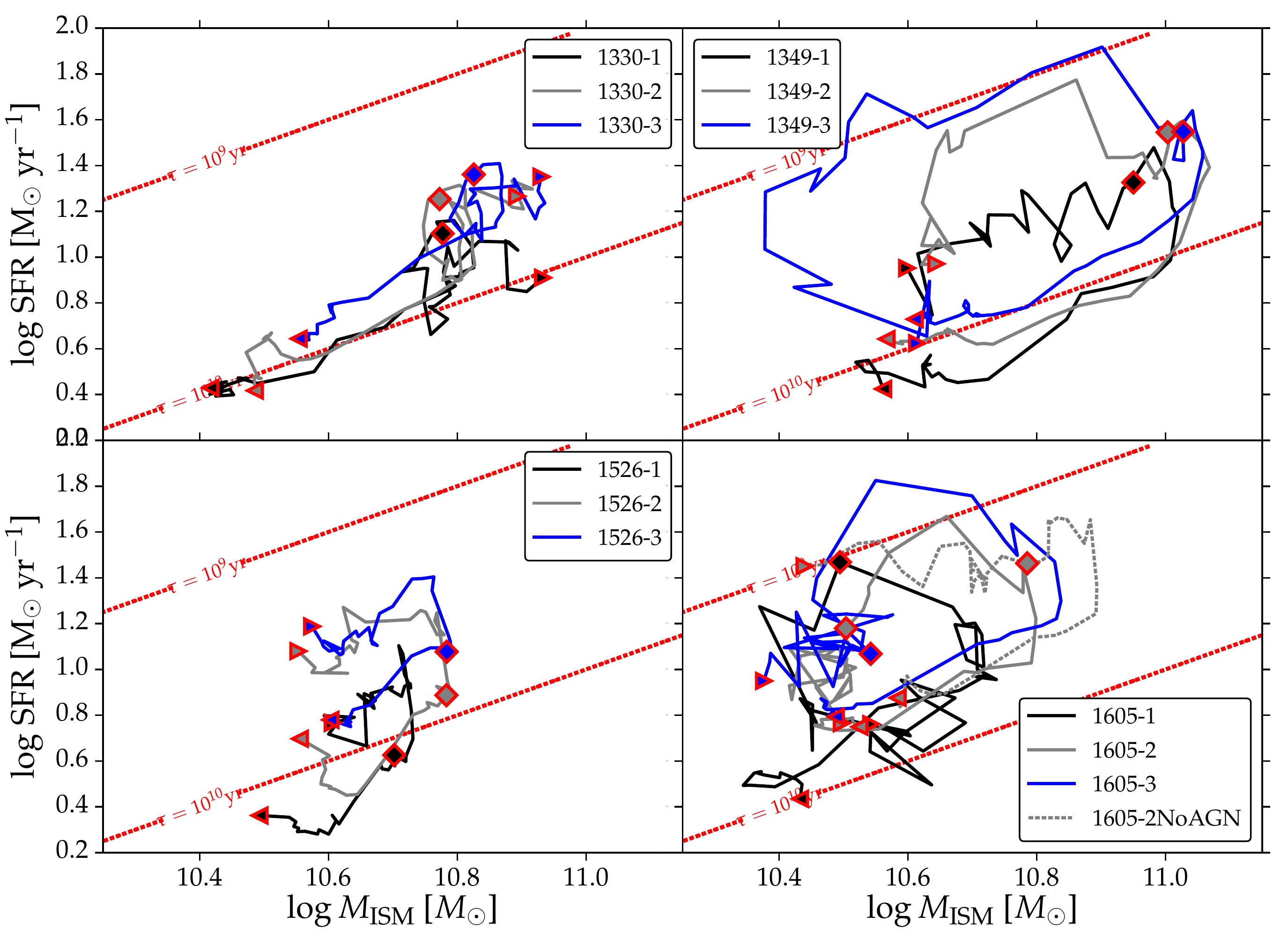} \caption{The SFR versus ISM gas mass between the beginning ($\triangleleft$-symbols) and end ($\triangleright$-symbols) of each merger. The merging time is marked with a $\Diamond$-symbol. Contours of constant ISM gas consumption timescale, $\tau\equiv M_\text{ISM}/\text{SFR}$, are shown by the \emph{dashed lines}. The other lines show the evolution of the simulated galaxies (with the same colour coding as in Figure~\ref{AllSFR}). SFR and $M_\text{ISM}$ are calculated based on the gas within a radius of 80 (physical) kpc of the galaxy at a given time. For simulations 1349, 1526 and 1605, the galaxies make a loop in this diagram. Step 1: ($\triangleleft$-symbols) before the merger the galaxies form stars with $\tau\simeq 10^{10}$ yr. Step 2 ($\triangleright$-symbols): when the cores of the two galaxies merge, the gas is compressed and a $\simeq 3-10$ times shorter gas consumption timescale is reached. This is also where the SFR peak appears. Step 3 ($\Diamond$-symbol, approximately): After coalescence of the cores, the SFR and amount of ISM gas decrease. For the 1330-X simulations the post-merger evolution is more complicated because several minor mergers occur. For the 1605-2 simulation we show a run without AGN feedback and black holes enabled (grey-dashed line).}
\label{PlotSFRVersusMISM}
\end{figure*}

Also shown in the figure is the stellar morphology (\emph{right panel}). The image has been constructed based on the $U$, $B$ and $K$ band luminosities (they represent \emph{blue}, \emph{green} and \emph{red}, respectively) of each stellar population particle belonging to the galaxy. Again, the colors represent the logarithm of the luminosity. The blue (young) stars show a ring- or disc-like morphology which closely traces the ISM gas. This is of course expected since stars are formed out of the ISM.  Another visible feature in the stellar distribution is the presence of a red bulge in a region with very little ISM gas.

\subsection{The SFR -- gas-mass plane}

The most important characteristic of a starburst galaxy is the timescale over which gas is consumed by star formation \citep[for an overview, see][]{2013seg..book..491S}. To reveal the presence of a bursty star formation mode in our simulations, we show SFR versus $M_\text{ISM}$ in Figure~\ref{PlotSFRVersusMISM}. Also shown are contours with constant gas depletion timescale, $\tau\equiv M_\text{ISM} / \text{SFR}$. In the literature, similar diagrams revealing the gas depletion timescale are often used to study the star formation laws of observed galaxies \citep{1991ApJ...370..158S,2014ApJ...793...19S,2015ApJ...800...20G, 2015arXiv151105149S}. The contours with a gas depletion timescale of 1 Gyr and 10 Gyr correspond to our bursty and normal mode, respectively. Of course the gas consumption timescales characteristic of the two modes depend on the definition of the gas mass. For our simulations the consumption timescale calculated based on the star-forming gas would for example be around four times shorter than for the total ISM gas. Similarly one would get a different timescale if only H$_2$ or HI gas was used. Some caution should thus be taken when comparing the timescales of our modes with observed consumption timescales.

In our analysis, the gas quantities are calculated by summing over all ISM gas cells within 80 (physical) kpc of the center of the main galaxy. This choice of radius is used to avoid a sudden transfer of mass between the two merging galaxies during their encounters, where the halo finder would classify the primary progenitor as a central galaxy, and the secondary galaxy as a satellite, and therefore momentarily associate a too large stellar mass with the primary galaxy (\citealt{2015MNRAS.449...49R}, see also \citealt{2015MNRAS.454.3020B}). A consequence of this definition is that our measurement of SFR and $M_\text{ISM}$ will contain all the gas within both galaxies even before they physically merge.

\begin{figure*}
\centering
\includegraphics[width = 0.9 \textwidth]{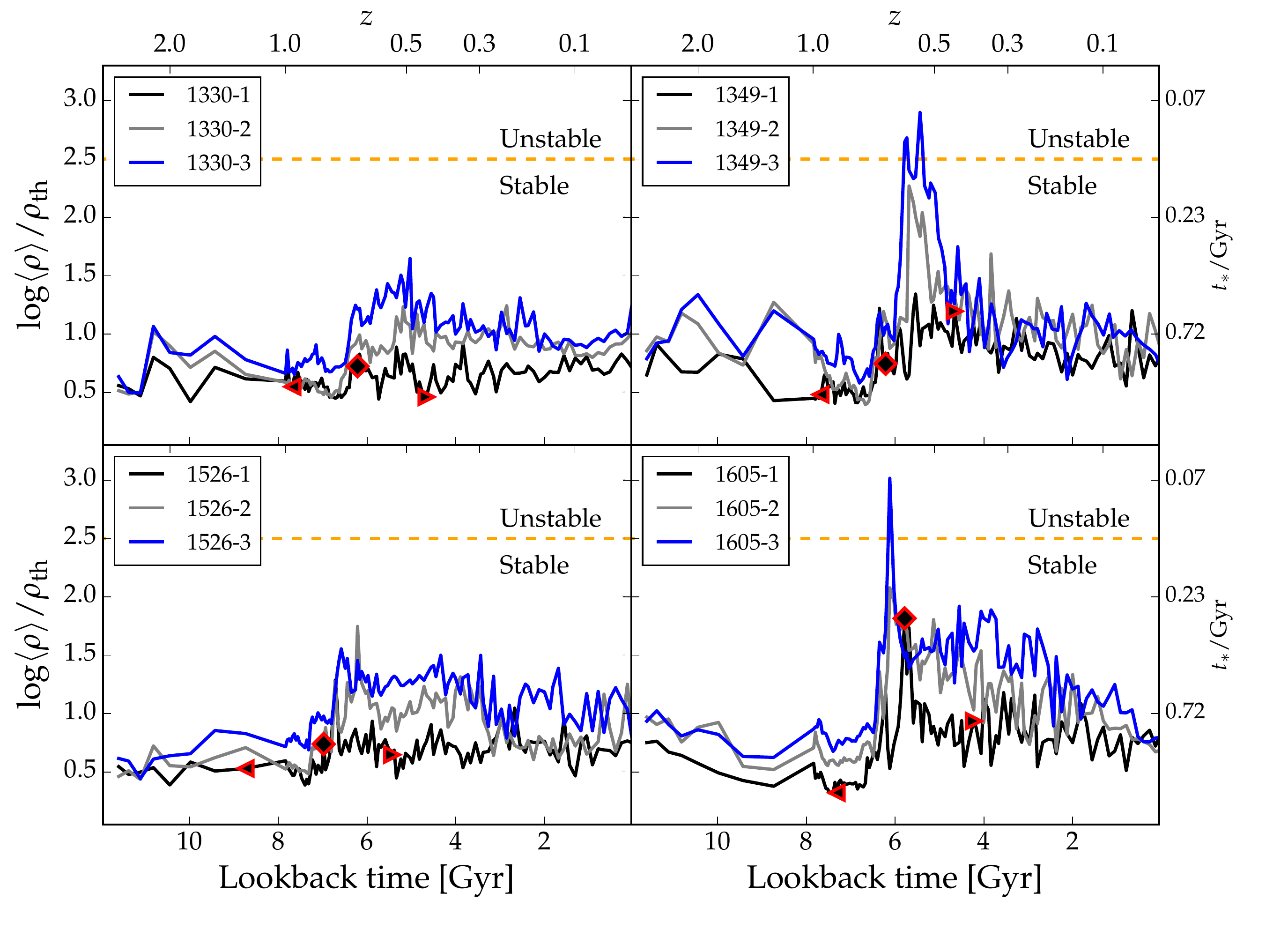}
\caption{The SFR-averaged density as function of lookback time for the simulated galaxies. The density is normalized to the SFR-threshold, $\rho_\text{th}$. Also indicated is the star-formation timescale (the second $y$-axis to the right of each plot). For the high-resolution simulations, star formation occurs at higher densities after the merger ($\gtrsim 1.5$ Gyr). In two of the galaxies, 1349-3 and 1605-3, the gas is compressed so much that the gas becomes unstable, causing stars being formed much more efficiently than in the normal star formation mode. }
\label{AllDensity.pdf}
\end{figure*}

During a merger, galaxies trace out a loop in the SFR--$M_\text{ISM}$ plane. This is for example clearly seen for galaxy 1349-2, which goes through the following three phases:
\begin{itemize}
\item[]\hspace*{-0.3cm}{\bf Step 1}: Before the merger the galaxy forms stars quiescently with a gas depletion timescale of $10^{10}$ yr, and an ISM gas mass of $10^{10.5} \msun$. Closer to the merger, new ISM gas enters the galaxy and the ISM mass increases to a value of $10^{11} \msun$, while the ISM depletion timescale remains constant. During this step the galaxy form stars at a gas depletion timescale characteristic for the \emph{normal star formation mode} of galaxies.
\item[]\hspace*{-0.3cm}{\bf Step 2}: Closer to the coalescence of the two galaxy cores, the galaxy evolves vertically in the SFR--$M_\text{ISM}$ diagram. Here the ISM gas mass remains constant, but an increased SFR appears because the gas depletion timescale becomes shorter. As we will see in Section~\ref{SFRRho}, the gas depletion timescale shortens because the gas is compressed to high densities.
\item[]\hspace*{-0.3cm}{\bf Step 3}: Immediately after the cores have merged, ISM gas is turned into stars at a short gas depletion timescale of $10^9-10^{9.5}$ yr. For some time after the merger the galaxy's gas depletion timescale remains low. The ISM gas mass declines at an approximately constant $\tau$-value of $10^{9.1}$ yr from $10^{10.9} \msun$ to $10^{10.6}\msun$. This constant $\tau$-value shows that for some time after the merger, the galaxy effectively has a memory of the merger event. At later times, the $\tau$-value increases again to a value close to (but slightly shorter than) the pre-merger value.
\end{itemize}

This picture is not only valid for galaxy 1349-2, but it also describes the processes happening in the 1349-X, 1526-X and 1605-X-simulations well. The presence of this evolutionary sequence shows that during a major merger, our galaxy formation models produce a phase of starbursting gas with $\simeq 10$ times shorter gas depletion timescales than the normal mode of star formation. This is the first time the presence of such starbursting gas is demonstrated in a cosmological simulation when an ISM subgrid model like ours is used. We note that a similar bursty star formation mode has previously been revealed by idealised merger simulations \citep{2010ApJ...720L.149T,2014MNRAS.442L..33R} performed with the RAMSES simulation code \citep{2002A&A...385..337T}. In Section~\ref{Discussion} we will further discuss implications of the starbursts and starbursting gas for hydrodynamical simulations.

The 1330-simulations show however a qualitatively different evolution than described by the above loop. The post-merger system is left with a larger amount of ISM gas than anytime during the merger. In Section~\ref{NoBurst}, we show that the reason for this is that the SFR-peak of the 1330-3 galaxy (and also for 1526-3) occurs when the two merging galaxies are several kpc apart, and the gas is therefore not compressed as much as in a nuclear starbursts.

In the \emph{lower right panel} we again compare the standard 1605-2 run to a simulation without black holes. The two runs behave qualitatively similar with a compression of gas to a shorter gas consumption timescale. The result that this merger leads to a starburst with a very short gas consumption timescale is therefore not significantly changed by the presence or absence of black hole feedback. A difference is, however, that the post-merger galaxy (marked with the $\triangleright$-symbol) for the run without black holes has a shorter gas consumption timescale than for the fiducial run. This is consistent with Section~\ref{SFHsubsection} where we found black holes to be important for shaping the post-merger evolution of the star formation history. It is possible that the detailed orbit of the black holes in a merger \citep[as recently modeled by][]{2015MNRAS.451.1868T} can potentially change the exact time, when strong feedback kicks in and heats the star-forming gas. We speculate that this is more likely to change the duration of a starburst epoch rather than changing whether starbursting gas appear in the galaxy or not.

\begin{figure*}
\centering
\includegraphics[width = 0.98 \textwidth]{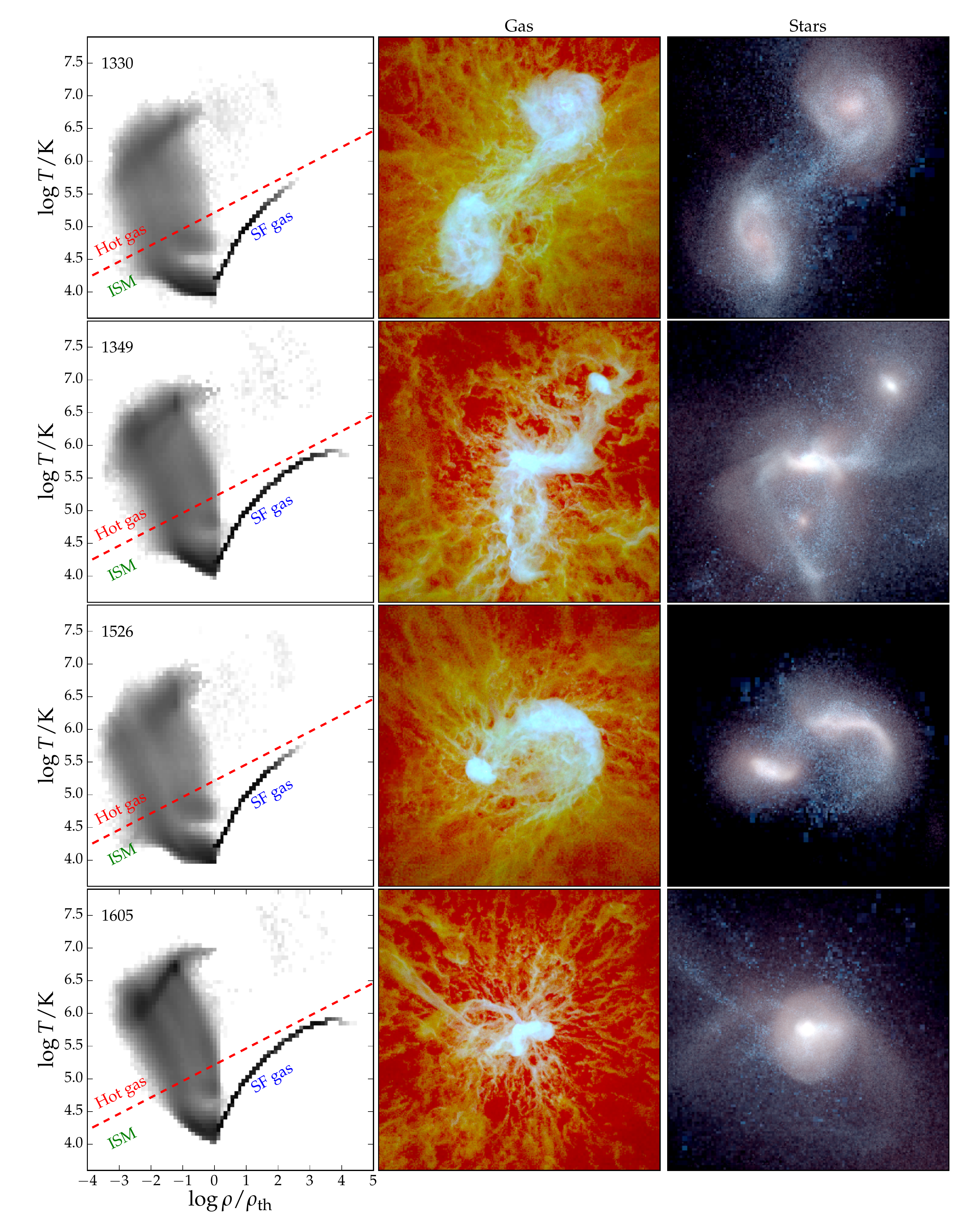}
\caption{The behaviour of the four simulations with zoom factor 3 at the time of the SFR peak. We show the density--temperature diagram (\emph{left panel}), the distribution of hot, ISM and star-forming gas (\emph{central panel}), and the stellar distribution (\emph{right panel}). The width of the images is 40 physical kpc. The color codings are the same as in Figure~\ref{RhoT}. In the 1330-3 and 1526-3 simulations (first and third row, respectively), the SFR peaks occur when the galaxies are far apart, and for the two other simulations the SFR peaks occur during a nuclear starburst. The gas of the two latter nuclear starburst galaxies also has a larger SFR peak because it is compressed to higher densities.}
\label{PlotDensityMapAtMaxSFR}
\end{figure*}

\subsection{The dense, starbursting gas}\label{SFRRho}

In our star formation model, the timescale on which stars are formed is closely tied to the local gas density. The time-evolution of the gas density at which stars are typically formed is further studied in Figure~\ref{AllDensity.pdf}, which shows the star-formation rate averaged gas density,
\begin{align}
\langle \rho \rangle = \frac{\sum_i \text{SFR}_i \rho_i}{\sum_i \text{SFR}_i}. \label{DensityDefinition}
\end{align}
Here the sum is over all gas cells within 80 kpc of a galaxy. In the figure, the density is normalized with respect to the density threshold, $\rho_\text{th}$, for star formation to occur. The SFR-weighting effectively makes the density-measurement more sensitive to the very dense gas with high SFRs.

For the 1330-X and 1526-X simulations, stars are typically formed at gas densities between 3 to 30 times (i.e.~0.5-1.5 dex above) the star formation threshold. The two other sets of simulations, 1349-X and 1605-X, show a resolution-dependent behaviour around the burst epoch. The low-resolution galaxies (1349-1 and 1605-1) lack very dense star-forming gas, but the simulations with higher resolution contain star-bursting gas with a much shorter star formation timescale. For the 1349-3 and 1605-3 simulations, the density at which stars are typically formed peaks at around $1000\rho_\text{th}$, which corresponds to a star formation timescale of 70 Myr.

At high densities of $\rho \gtrsim 10^{2.5}\rho_\text{crit}$, the gas in the multi-phase ISM model has properties similar to starbursting gas. At such densities the effective equation of state describing the star-forming gas has a polytropic index, $\text{d}\ln P/\text{d}\ln \rho$, smaller than $4/3$, implying that gravitational instabilities set in. Furthermore, above the same density threshold, the star formation timescale is shorter than the cloud evaporation timescale implying that a large fraction of the mass contained in a gas cell is converted into stars before the cold clouds are evaporated by thermal conduction. When the gas reaches such high densities it will thus form stars much more efficiently than normal star-forming gas. This instability threshold is marked by the \emph{dashed horizontal line} in Figure~\ref{AllDensity.pdf}, and it is seen that only the 1349-3 and 1605-3 simulations reach average densities above this threshold.

It is also worthwhile to point out that an increase in the density at which stars are formed is not only restricted to a short time-interval around the peak of the star formation rate, when the cores of the two galaxies merge. For all the high-resolution simulations, the SFR-averaged density is elevated by about 0.8-1 dex for several Gyr after the merger (compared to the value before the merger). Thus, a merger does not only affect the star formation properties of the gas during the coalescence of the cores, but it also impacts the gas in the post-starburst phase. For the lower resolution run, the gas density at which stars are formed quickly loses its memory about the merger, and shortly after the SFR-peak stars are again formed at similar densities as for the pre-merger galaxy. Such a bursty epoch of star formation (which is not necessarily coincident in time with a merger) is also favored by several recent observations of burst cycles in galaxies \citep{2012ApJ...744...44W,2014MNRAS.441.2717K,2016arXiv160405314G}.

The presence of very dense starbursting gas in the 1349-3 and 1605-3 simulations is also reflected in the peak SFR of $70-100 \,\msun \,\text{yr}^{-1}$ revealed by Figure~\ref{AllSFR}. This is more than a factor of two higher than for the two high-resolution simulations, 1330-3 and 1526-3, that do not contain dense starbursting gas. Similarly, the 1349-3 and 1605-3 galaxies also reach a shorter gas depletion timescale during the merger according to the SFR--$M_\text{ISM}$-plane in Figure~\ref{PlotSFRVersusMISM}.

In Section~\ref{SFHsubsection} we noted that the star formation histories of galaxies are not perfectly converged at redshifts $1\leq z\leq 2$. In Figure~\ref{AllDensity.pdf}, we see that star formation typically occurs in more dense regions, when the resolution of a simulation is increased. This trend is especially clear at $1\leq z\leq 2$. This behaviour -- that star formation occurs at higher densities when the resolution is increased -- is the reason for the increase in the SFR at increased resolution. We note that the lines in Figure~\ref{AllDensity.pdf} are not so well sampled at high redshift because frequent snapshots were only stored at $z<1$, which is the primary epoch of interest for our merger studies.

\begin{figure*}
\centering
\includegraphics[width = 0.9 \textwidth]{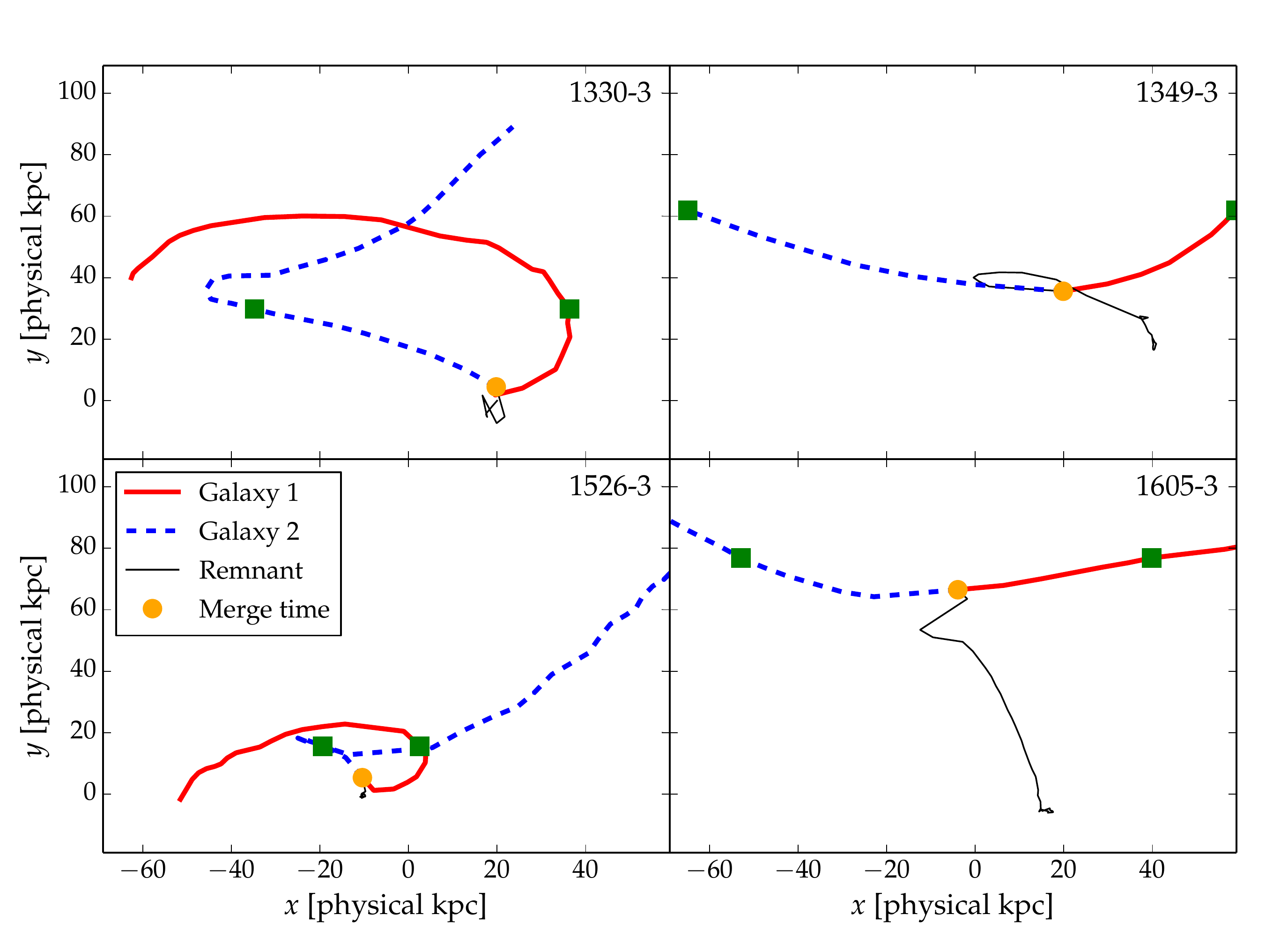}
\caption{Trajectories of the galaxies for the high-resolution simulations in a $x-y$ projection in a coordinate system where the $z$-axis is perpendicular to the orbital plane. The \emph{thick} (\emph{red}) \emph{line} shows the galaxy with the most massive stellar component at $z=0.93$, and the \emph{dashed} (\emph{blue}) \emph{line} shows the other galaxy participating in the merger. The trajectory of the merger remnant is shown as a \emph{thin black line}. The orbital plane is chosen such that it contains the merger remnant immediately after the merger (\emph{orange circle}), and the position of each galaxy well before the merger (\emph{green squares}, see text for details). Simulations 1349-3 and 1605-3 are head-on mergers, whereas the galaxies in the two other mergers are orbiting each other before they coalesce. Interestingly, only the head-on mergers go through a nuclear starburst phase (see Figures~\ref{AllDensity.pdf}~and~\ref{PlotDensityMapAtMaxSFR}).}
\label{Orbits1_plane_projection}
\end{figure*}

\subsection{Morphology of the star-forming gas}

To further examine the behavior of the starbursting gas in the major mergers, we will analyse the morphology of the gas and stars in our galaxies. For each high-resolution simulation with a zoom-factor of 3 we identify the time of the merger induced SFR peak. We simply select this time as the snapshot with the highest SFR at a lookback time larger than 5 Gyr. For each of these snapshots we plot the density--temperature diagram and the spatial distribution of gas and stars (see Figure~\ref{PlotDensityMapAtMaxSFR}). All these characteristics have already been introduced in Section~\ref{ISMExplanation}.

\subsubsection{The two galaxies with strong nuclear starbursts (1349-3 and 1605-3)}

The density--temperature diagram for the galaxies reveals that very dense star-forming gas with $\rho\gtrsim 10^{3.5} \rho_\text{th}$ only exists for two of the four major mergers. These two galaxies (1349-3 and 1605-3) are, of course, the same galaxies that were found to form stars at high densities in Figure~\ref{AllDensity.pdf}. The stellar morphology reveals that these two galaxies have their SFR peak at the time when the cores of the colliding galaxies merge. The 1605-3 simulation reveals an extremely dense stellar morphology. The 1349-3 simulation also shows two galaxies with merging cores, and in addition, several other galaxies are also visible, and a number of streams mainly caused by the merger. The gas morphology also shows signs of a nuclear starburst for these systems with a single very dense region with a large amount of star-forming gas (see \emph{white--blue colors} in the \emph{central panels} in \emph{row 2 and 4}).

\subsubsection{The two galaxies with a SFR peak at a 10-30 kpc separation (1330-3 and 1526-3)}\label{NoBurst}

The two mergers, which do not contain dense star-bursting gas in Figure~\ref{PlotDensityMapAtMaxSFR} have their SFR peak when their separations are around 10-30 kpc according to the stellar light distribution in the \emph{right panels} (for simulation 1349-3 and 1605-3). Again, the star-forming gas closely traces the morphology of the recently formed stars. The projected gas distributions reveal that the star-forming region of these two galaxies is distributed over a larger area for the nuclear starbursts. A larger spatial extent of the gas is consistent with the gas being less compressed. Therefore, also the SFR enhancements at the peak are smaller for these galaxies, as revealed by Figure~\ref{AllSFR}.

\begin{figure*} \centering \includegraphics[width = 0.9 \textwidth]{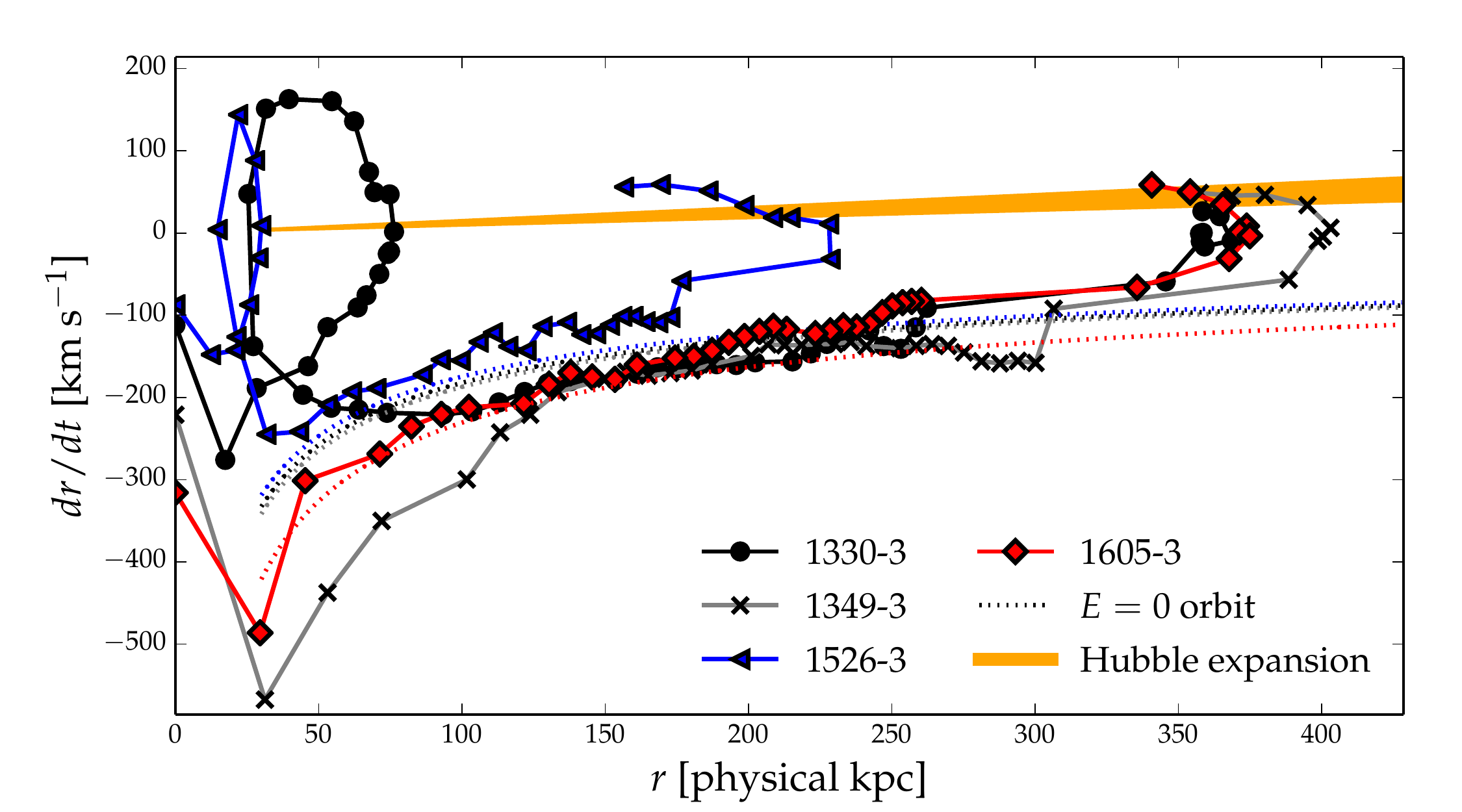} \caption{Each merger orbit characterised by the physical distance ($r$) between the two galaxies participating in the merger, and the radial velocity, ${\rm d}r/{\rm d}t$. Orbits are shown for each of the high-resolution simulations (\emph{circles, triangles, diamonds} and \emph{crosses}). The halo masses are measured at $z=0.93$, and for each system we have calculated the expected evolution of galaxies in the ($r$, ${\rm d}r/{\rm d}t$)-plane for an $E=0$ orbit. The shaded (\emph{orange}) area is bounded by lines showing the Hubble flow at $z=0.5$ and $z=1.5$ (120 and 158 km s$^{-1}$ Mpc$^{-1}$, respectively), and is thus showing the effect of Hubble expansion at the epoch relevant for the merger orbits. The galaxies with intense nuclear starbursts (1349-3 and 1605-3) approach each other faster than the two other galaxies (when measuring ${\rm d}r/{\rm d}t$ at, e.g., 30 kpc).}
\label{Orbits}
\end{figure*}

\section{Orbit of the mergers}\label{OrbitSection}

It is natural to ask why some major mergers end up as nuclear starbursts, and others do not. A key to understand this lies in the orbital parameters of the galaxies, which we study in this section. Most of the existing literature about mergers relies on a range of assumptions about the initial orbits of the galaxies participating in the merger. Typically, galaxies are assumed to be on a Keplerian orbit with an energy
\begin{align}
E = \frac{1}{2} \mu \left(\frac{{\rm d}r}{{\rm d}t}\right)^2+\frac{1}{2}\frac{L^2}{\mu r^2}-\frac{GM_\text{halo,1}M_\text{halo,2}}{r},\label{phasespace}
\end{align}
where $\mu$ is the reduced mass of the system,
\begin{align}
\mu \equiv \frac{M_\text{halo,1}M_\text{halo,2}}{M_\text{halo,1}+M_\text{halo,2}},
\end{align}
and $M_\text{halo,1}$ and $M_\text{halo,2}$ are the masses of the two galaxies participating in the merger. $r$ is the distance between them, and $L$ is the angular momentum of the orbit. Elliptical, parabolic and hyperbolic orbits have $E<0$, $E=0$ and $E>0$, respectively.

We will now examine how well the orbits of our cosmological mergers fit with such Keplerian orbits. We will track the positions in physical coordinates of the two merging galaxies and the merger remnant in each simulation. We do this in the rest frame of the $z=0$ merger remnant. To determine the orbital plane, it is necessary to know three points within the plane. As the first two points we will use the positions of our galaxies before the merger at $z=0.75$ (for 1330-3 we use $z=0.6$, because the merger happens slightly later than for the other systems). As the third point we use the position of the merger remnant in the first snapshot after the merger occurred.

Figure~\ref{Orbits1_plane_projection} shows the galaxy trajectories projected onto the orbital plane. Here the different lines connect the positions of galaxies of our snapshots. The coordinates used to determine the orbital planes are shown as \emph{circles} (the merger remnant's position) and \emph{squares} (the positions before the merger; $z=0.6$ for 1330-3 and $z=0.75$ for the other galaxies). The orbits for the 1349-3 and 1605-3 galaxies are clearly head-on mergers, where the galaxies do not circulate each other before they coalesce. The two other mergers, 1330-3 and 1526-3, show a different behaviour, where the galaxies pass each other before they merge. The two head-on mergers are the systems that lead to nuclear starbursts with extremely dense gas (see section~\ref{SFRRho}), whereas the other two galaxies do not exhibit strong bursts.

\subsection{Energy of the orbits}

Keplerian orbits follow a path in the ($r$, ${\rm d}r/{\rm d}t$) phase-space diagram predicted by Eqn.~(\ref{phasespace}). For a fixed energy $E$ and angular momentum $L$, it is therefore straightforward to calculate the evolution for the Keplerian prediction. To quantify how strongly galaxies deviate from simple Keplerian orbits we plot the corresponding phase-space diagram in Figure~\ref{Orbits}. For each snapshot of each merger, we calculate the (physical) distance between the two merging galaxies as a function of time. ${\rm d}r/{\rm d}t$ is obtained by a simple numerical differentiation.

We compare our orbits with a head-on orbit with $L=0$ and $E=0$. The speed at which the two galaxies approach each other for such an orbit is the highest possible if the two galaxies participating in the merger were not accelerated by other sources than each other. We fix the halo mass of each system to the $z=0.93$ value.  Interestingly, the 1349-3 is accelerated to a velocity exceeding the $E=0$ energy (i.e.~above the escape velocity in Newtonian mechanics). This galaxy experiences additional gravitational kicks from galaxies other than the ones participating in the merger. In fact, in Figure~\ref{PlotDensityMapAtMaxSFR} we saw that several galaxies participate in the merger of the 1349-simulation. It is an important result that in cosmological setups colliding galaxies can approach each other faster than predicted by $E=0$ orbits because of additional velocity kicks from the environment in which the merger takes place.

The presence of gravitational kicks beyond the interaction between the merging galaxies is also visible for the 1330-3 simulation, when the galaxies are 260 kpc apart. The orbit of the 1605-3 merger also shows evidence for additional accelerations since the galaxies approach each other at a speed smaller than the $E=0$ orbit until they are at a 150-250 kpc separation, at which point they are accelerated to a speed similar to the shown trajectory with $E=0$. The 1330-3 and 1526-3 mergers approach each other at a speed smaller than the $E=0$ orbit. 

Also visible from the figure is that the mergers that lead to starbursts are those which approach each other at the highest speed. For 1349-3 and 1605-3, the galaxies move toward each other with 480 and 320 km s$^{-1}$, respectively. The two characteristics that appear special for simulations with nuclear starbursts are that, (1) the collisions are head-on, and (2) the collisions occur at higher velocities than for the galaxies without a nuclear starburst.

\section{Discussion: starbursts in galaxy simulations} \label{Discussion}

A characteristic feature of the ISM model adopted in this paper is that the effective pressure of the star-forming gas increases faster with density than for an isothermal gas \citep[see figure 4 in][]{2005MNRAS.361..776S}. This increased pressure accounts for feedback processes related to star formation. The approach of not resolving the small-scale structure of the ISM explicitly and instead imposing an effective equation of state governing the star-forming gas has been adopted in the Illustris simulation as well. This has helped to make such cosmological hydrodynamical simulations of galaxy formation feasible, and in turn allowed them to contribute to our understanding of the formation of spiral and elliptical galaxies \citep{2014Natur.509..177V}, the formation of compact quiescent galaxies at $z=2$ \citep{2015MNRAS.449..361W}, the colours of satellite galaxies \citep{2015MNRAS.447L...6S}, and how feedback affects gas accretion and the circumgalactic medium of galaxies (\citealt{2015MNRAS.448...59N} and \citealt{2015MNRAS.448..895S}, respectively). A similar -- although slightly different in detail -- prescription of the ISM gas has successfully reproduced many galaxy observables in the large-scale cosmological simulation EAGLE \citep{2015MNRAS.446..521S,2015MNRAS.450.1937C,2015MNRAS.451.1247S,2015MNRAS.450.4486F}.  But since both of these approaches do not treat the small-scale structure of the ISM realistically, it is expected that the short-time variability in the SFR is generally not completely accounted for.

In this study, we have for the first time produced realistic starburst galaxies within such subgrid frameworks using full cosmological simulations. We stress that it is an important result of any ISM model to be able to produce the enhanced SFRs and short gas depletion timescales as observed in starburst galaxies. We have here found that this starbursting phase occurs in major mergers. The compression of gas caused by large-scale gas flows when two galaxies merge is the key driver of the starburst events. The gas flows succeed in driving starbursts even though we use a heavily simplified ISM model that normally does not admit large density fluctuations in the ISM. Since we have only considered major mergers here, it is in principle possible that other environmental effects such as minor mergers or filamentary accretion \citep{2005MNRAS.363....2K,2009Natur.457..451D,2013MNRAS.429.3353N} can also produce starbursting gas. Understanding this will be an interesting subject of future studies.

In our low resolution simulations (with a `zoom factor' of 1), the bursty star formation mode with short gas depletion times is essentially absent. These simulations have a similar resolution as Illustris, and it is therefore not surprising that Illustris lacks a population of extreme starburst galaxies. With our zoom simulations of major mergers we have shown that increasing the mass-resolution of the simulations by a factor of 10 to 40 is all what is required for bursty star formation to appear. This finding is consistent with the speculation in \citet{2015MNRAS.447.3548S}, where it was suggested that the reason for the dearth of starbursts in Illustris is a too low resolution that is insufficient to resolve the sub-kpc starbursting regions.

A completely different approach to form starbursts in hydrodynamical simulations would be to use a different model of the galaxy ISM. \citet{2014MNRAS.445..581H} for example use a feedback model that (attempts to) describe star-forming clouds inside galaxies. This creates rapidly fluctuating star formation rates in dwarf galaxies \citep{2015arXiv151003869S}, which can account for the observed H$\alpha$ and far-UV fluxes of local galaxies \citep{2012ApJ...744...44W}. Also, the ISM has a more clumpy morphology and a larger star formation rate threshold, which also increase the burstiness of these simulations. Other examples of more bursty star formation models are \citet{2013MNRAS.428..129S} and \citet{2010Natur.463..203G}. In these models stars are usually formed in sub-kpc sized clumps of star-forming gas. We have here shown that it is not strictly necessary to adopt such clumpy ISM models in order to produce nuclear starbursts; the global gas flows in our major mergers in combination with a subgrid prescription of the ISM and high resolution are sufficient.

\section{Conclusions} \label{Conclusion}

In this paper, we have studied gas properties in a set of high-resolution cosmological zoom simulations of major mergers. The mergers occur over the redshift range $0.5<z<1$, and the participating galaxies have masses in the range $10^{9.65}\leq M_*/\msun\leq 10^{10.27}$. With our numerical setup we have re-simulated individual galaxies from Illustris with an up to 40 times better mass resolution, but still including a physical treatment of the full cosmological environment surrounding the galaxies of interest. In the present work we have focused on the gas and starburst properties of the major mergers, and our main results are:

\begin{itemize}
\item High resolution merger simulations reveal a phase of starbursting gas with $\simeq$10 times shorter ISM gas depletion timescales than normal star-forming galaxies. This bursty mode of star formation appears when the cores of two merging galaxies coalesce. A characteristic of the bursty mode is that the star-forming gas is so dense that the ISM becomes unstable. The bursts in our simulations are similar to those observed in ULIRGS \citep{1991ApJ...370..158S}. This is the first time such a bursty star formation mode has been revealed in full cosmological simulations within the framework of the subgrid ISM models.
\item Only two out of four major mergers we studied go through a nuclear starburst episode. The two other galaxies have their SFRs peak when the two galaxies are still 10-30 kpc apart. Our cosmological simulations therefore predict that only a fraction of all gas-rich major mergers go through an intensive nuclear starburst phase. We find that the major mergers leading to strong nuclear starbursts are head-on mergers, whereas collisions with large impact parameters have their SFR peaks when the galaxies are 10-30 kpc apart. A characteristic of the nuclear starbursts is also that the colliding galaxies have a larger collision velocity than the other systems. In one of the cases, the galaxies collide at a speed two times larger than predicted by the $E=0$ parabolic orbit. We note, however, that these conclusions are only based on a sample of four mergers. Assembling larger samples is clearly desirable to gain better statistics for the behaviour of starburst galaxies in cosmological simulations.
\item We argue that it is an important finding that the simple subgrid star formation model, as applied for example in Illustris, can produce realistic starbursts with two modes of star formation distinguished by their ISM gas depletion timescale. Using such a simple star formation model therefore does not preclude the occurrence of extreme starbursts. This starbursting mode is, however, only fully revealed at a 10-40 times higher mass resolution than used in Illustris. This is still lower than the resolution requirements of simulations that attempt to resolve the ISM explicitly. The main barrier for reproducing 
the number of starbursting galaxies in Illustris thus not appears to lie in limitations of the physics model, but rather in the limited resolution of the simulations \citep[see also][]{2015MNRAS.447.3548S}.
\end{itemize}

Most of the literature about major merger simulations relies on idealised set-ups where equilibrium galaxies are collided in isolation (an exception is \citealt{2009MNRAS.398..312G}).  This involves a large set of ad-hoc assumptions about the structural properties of the involved galaxies and the orbital parameters of the encounter. In this simulation suite we have taken a different approach and simulated four different major mergers consistently embedded in a $\Lambda$CDM universe by means of the zoom technique. This thus opens up the chance to address the physical characteristics of mergers in a realistic cosmological environment, eliminating the parameter freedom that troubles isolated toy simulations. In the future, we plan to further apply this powerful approach to study the formation, destruction and re-growth of stellar discs before, during and after the occurrence of mergers, respectively. A related point will be to see how the $z=0$ stellar distribution are affected by mergers at $z=0.5-1$.

\section*{Acknowledgements}

We thank the referee for constructive and useful comments. MS thanks the Sapere Aude fellowship program. VS acknowledges the European Research Council through ERC-StG grant EXAGAL-308037, and the SFB-881 `The Milky Way System' of the German Science Foundation. The Dark Cosmology Centre is funded by the Danish National Research Foundation. The authors like to thank the Klaus Tschira Foundation.

\def\aj{AJ}
\def\araa{ARA\&A}
\def\apj{ApJ}
\def\apjl{ApJ}
\def\apjs{ApJS}
\def\apss{Ap\&SS}
\def\aap{A\&A}
\def\aapr{A\&A~Rev.}
\def\aaps{A\&AS}
\def\mnras{MNRAS}
\def\nat{Nature}
\def\pasp{PASP}
\def\aplett{Astrophys.~Lett.}
\def\physrep{Physical Reviews}
\def\nar{New A Rev.}

\footnotesize{
\bibliographystyle{mn2e}
\bibliography{ref}

\begin{thebibliography}{69}
\expandafter\ifx\csname natexlab\endcsname\relax\def\natexlab#1{#1}\fi

\bibitem[{{Bauer} \& {Springel}(2012)}]{2012MNRAS.423.2558B}
{Bauer} A., {Springel} V., 2012, \mnras, 423, 2558

\bibitem[{{Behroozi} {et~al}\mbox{.}(2015){Behroozi}, {Knebe}, {Pearce},
  {Elahi}, {Han}, {Lux}, {Mao}, {Muldrew}, {Potter}, \&
  {Srisawat}}]{2015MNRAS.454.3020B}
{Behroozi} P. {et~al.}, 2015, \mnras, 454, 3020

\bibitem[{{Bigiel} {et~al}\mbox{.}(2011){Bigiel}
  {et~al.}}]{2011ApJ...730L..13B}
{Bigiel} F., {et~al.}, 2011, \apjl, 730, L13

\bibitem[{{Brinchmann} {et~al}\mbox{.}(2004){Brinchmann}, {Charlot}, {White},
  {Tremonti}, {Kauffmann}, {Heckman}, \& {Brinkmann}}]{2004MNRAS.351.1151B}
{Brinchmann} J., {Charlot} S., {White} S.~D.~M., {Tremonti} C., {Kauffmann} G.,
  {Heckman} T., {Brinkmann} J., 2004, \mnras, 351, 1151

\bibitem[{{Chabrier}(2003)}]{2003PASP..115..763C}
{Chabrier} G., 2003, \pasp, 115, 763

\bibitem[{{Cox} {et~al}\mbox{.}(2006){Cox}, {Jonsson}, {Primack}, \&
  {Somerville}}]{2006MNRAS.373.1013C}
{Cox} T.~J., {Jonsson} P., {Primack} J.~R., {Somerville} R.~S., 2006, \mnras,
  373, 1013

\bibitem[{{Crain} {et~al}\mbox{.}(2015){Crain}, {Schaye}, {Bower}, {Furlong},
  {Schaller}, {Theuns}, {Dalla Vecchia}, {Frenk}, {McCarthy}, {Helly},
  {Jenkins}, {Rosas-Guevara}, {White}, \& {Trayford}}]{2015MNRAS.450.1937C}
{Crain} R.~A. {et~al.}, 2015, \mnras, 450, 1937

\bibitem[{{Dav{\'e}} {et~al}\mbox{.}(2011){Dav{\'e}}, {Oppenheimer}, \&
  {Finlator}}]{2011MNRAS.415...11D}
{Dav{\'e}} R., {Oppenheimer} B.~D., {Finlator} K., 2011, \mnras, 415, 11

\bibitem[{{Dekel} {et~al}\mbox{.}(2009){Dekel}, {Birnboim}, {Engel},
  {Freundlich}, {Goerdt}, {Mumcuoglu}, {Neistein}, {Pichon}, {Teyssier}, \&
  {Zinger}}]{2009Natur.457..451D}
{Dekel} A. {et~al.}, 2009, \nat, 457, 451

\bibitem[{{Faucher-Gigu{\`e}re} {et~al}\mbox{.}(2009){Faucher-Gigu{\`e}re},
  {Lidz}, {Zaldarriaga}, \& {Hernquist}}]{2009ApJ...703.1416F}
{Faucher-Gigu{\`e}re} C.-A., {Lidz} A., {Zaldarriaga} M., {Hernquist} L., 2009,
  \apj, 703, 1416

\bibitem[{{Furlong} {et~al}\mbox{.}(2015){Furlong}, {Bower}, {Theuns},
  {Schaye}, {Crain}, {Schaller}, {Dalla Vecchia}, {Frenk}, {McCarthy}, {Helly},
  {Jenkins}, \& {Rosas-Guevara}}]{2015MNRAS.450.4486F}
{Furlong} M. {et~al.}, 2015, \mnras, 450, 4486

\bibitem[{{Genel} {et~al}\mbox{.}(2014){Genel}, {Vogelsberger}, {Springel},
  {Sijacki}, {Nelson}, {Snyder}, {Rodriguez-Gomez}, {Torrey}, \&
  {Hernquist}}]{2014MNRAS.445..175G}
{Genel} S. {et~al.}, 2014, \mnras, 445, 175

\bibitem[{{Genzel} {et~al}\mbox{.}(2015){Genzel}, {Tacconi}, {Lutz},
  {Saintonge}, {Berta}, {Magnelli}, {Combes}, {Garc{\'{\i}}a-Burillo}, {Neri},
  {Bolatto}, {Contini}, {Lilly}, {Boissier}, {Boone}, {Bouch{\'e}}, {Bournaud},
  {Burkert}, {Carollo}, {Colina}, {Cooper}, {Cox}, {Feruglio}, {F{\"o}rster
  Schreiber}, {Freundlich}, {Gracia-Carpio}, {Juneau}, {Kovac}, {Lippa},
  {Naab}, {Salome}, {Renzini}, {Sternberg}, {Walter}, {Weiner}, {Weiss}, \&
  {Wuyts}}]{2015ApJ...800...20G}
{Genzel} R. {et~al.}, 2015, \apj, 800, 20

\bibitem[{{Governato} {et~al}\mbox{.}(2010){Governato}, {Brook}, {Mayer},
  {Brooks}, {Rhee}, {Wadsley}, {Jonsson}, {Willman}, {Stinson}, {Quinn}, \&
  {Madau}}]{2010Natur.463..203G}
{Governato} F. {et~al.}, 2010, \nat, 463, 203

\bibitem[{{Governato} {et~al}\mbox{.}(2009){Governato}, {Brook}, {Brooks},
  {Mayer}, {Willman}, {Jonsson}, {Stilp}, {Pope}, {Christensen}, {Wadsley}, \&
  {Quinn}}]{2009MNRAS.398..312G}
{Governato} F. {et~al.}, 2009, \mnras, 398, 312

\bibitem[{{Grand} {et~al}\mbox{.}(2016){Grand}, {Springel}, {G{\'o}mez},
  {Marinacci}, {Pakmor}, {Campbell}, \& {Jenkins}}]{2015arXiv151202219G}
{Grand} R.~J.~J., {Springel} V., {G{\'o}mez} F.~A., {Marinacci} F., {Pakmor}
  R., {Campbell} D.~J.~R., {Jenkins} A., 2016, \mnras, 459, 199

\bibitem[{{Guedes} {et~al}\mbox{.}(2011){Guedes}, {Callegari}, {Madau}, \&
  {Mayer}}]{2011ApJ...742...76G}
{Guedes} J., {Callegari} S., {Madau} P., {Mayer} L., 2011, \apj, 742, 76

\bibitem[{{Guo} {et~al}\mbox{.}(2016){Guo}, {Rafelski}, {Faber}, {Koo},
  {Krumholz}, {Trump}, {Willner}, {Amor{\'{\i}}n}, {Barro}, {Bell}, {Gardner},
  {Gawiser}, {Hathi}, {Koekemoer}, {Pacifici}, {P{\'e}rez-Gonz{\'a}lez},
  {Ravindranath}, {Reddy}, {Teplitz}, \& {Yesuf}}]{2016arXiv160405314G}
{Guo} Y. {et~al.}, 2016, ArXiv: 1604.05314

\bibitem[{{Hayward} {et~al}\mbox{.}(2014){Hayward}, {Torrey}, {Springel},
  {Hernquist}, \& {Vogelsberger}}]{2014MNRAS.442.1992H}
{Hayward} C.~C., {Torrey} P., {Springel} V., {Hernquist} L., {Vogelsberger} M.,
  2014, \mnras, 442, 1992

\bibitem[{{Hopkins} {et~al}\mbox{.}(2014){Hopkins}, {Kere{\v s}}, {O{\~n}orbe},
  {Faucher-Gigu{\`e}re}, {Quataert}, {Murray}, \&
  {Bullock}}]{2014MNRAS.445..581H}
{Hopkins} P.~F., {Kere{\v s}} D., {O{\~n}orbe} J., {Faucher-Gigu{\`e}re} C.-A.,
  {Quataert} E., {Murray} N., {Bullock} J.~S., 2014, \mnras, 445, 581

\bibitem[{{Kauffmann}(2014)}]{2014MNRAS.441.2717K}
{Kauffmann} G., 2014, \mnras, 441, 2717

\bibitem[{{Kennicutt}(1998)}]{1998ApJ...498..541K}
{Kennicutt}, Jr. R.~C., 1998, \apj, 498, 541

\bibitem[{{Kere{\v s}} {et~al}\mbox{.}(2005){Kere{\v s}}, {Katz}, {Weinberg},
  \& {Dav{\'e}}}]{2005MNRAS.363....2K}
{Kere{\v s}} D., {Katz} N., {Weinberg} D.~H., {Dav{\'e}} R., 2005, \mnras, 363,
  2

\bibitem[{{Knapen} \& {James}(2009)}]{2009ApJ...698.1437K}
{Knapen} J.~H., {James} P.~A., 2009, \apj, 698, 1437

\bibitem[{{Krumholz} {et~al}\mbox{.}(2012){Krumholz}, {Dekel}, \&
  {McKee}}]{2012ApJ...745...69K}
{Krumholz} M.~R., {Dekel} A., {McKee} C.~F., 2012, \apj, 745, 69

\bibitem[{{Marinacci} {et~al}\mbox{.}(2014){Marinacci}, {Pakmor}, \&
  {Springel}}]{2014MNRAS.437.1750M}
{Marinacci} F., {Pakmor} R., {Springel} V., 2014, \mnras, 437, 1750

\bibitem[{{McMillan}(2011)}]{2011MNRAS.414.2446M}
{McMillan} P.~J., 2011, \mnras, 414, 2446

\bibitem[{{Mihos} \& {Hernquist}(1994)}]{1994ApJ...431L...9M}
{Mihos} J.~C., {Hernquist} L., 1994, \apjl, 431, L9

\bibitem[{{Mihos} \& {Hernquist}(1996)}]{1996ApJ...464..641M}
{Mihos} J.~C., {Hernquist} L., 1996, \apj, 464, 641

\bibitem[{{Narayanan} {et~al}\mbox{.}(2010){Narayanan}, {Hayward}, {Cox},
  {Hernquist}, {Jonsson}, {Younger}, \& {Groves}}]{2010MNRAS.401.1613N}
{Narayanan} D., {Hayward} C.~C., {Cox} T.~J., {Hernquist} L., {Jonsson} P.,
  {Younger} J.~D., {Groves} B., 2010, \mnras, 401, 1613

\bibitem[{{Nelson} {et~al}\mbox{.}(2015{\natexlab{a}}){Nelson}, {Genel},
  {Vogelsberger}, {Springel}, {Sijacki}, {Torrey}, \&
  {Hernquist}}]{2015MNRAS.448...59N}
{Nelson} D., {Genel} S., {Vogelsberger} M., {Springel} V., {Sijacki} D.,
  {Torrey} P., {Hernquist} L., 2015{\natexlab{a}}, \mnras, 448, 59

\bibitem[{{Nelson} {et~al}\mbox{.}(2015{\natexlab{b}}){Nelson}, {Pillepich},
  {Genel}, {Vogelsberger}, {Springel}, {Torrey}, {Rodriguez-Gomez}, {Sijacki},
  {Snyder}, {Griffen}, {Marinacci}, {Blecha}, {Sales}, {Xu}, \&
  {Hernquist}}]{2015A&C....13...12N}
{Nelson} D. {et~al.}, 2015{\natexlab{b}}, Astronomy and Computing, 13, 12

\bibitem[{{Nelson} {et~al}\mbox{.}(2013){Nelson}, {Vogelsberger}, {Genel},
  {Sijacki}, {Kere{\v s}}, {Springel}, \& {Hernquist}}]{2013MNRAS.429.3353N}
{Nelson} D., {Vogelsberger} M., {Genel} S., {Sijacki} D., {Kere{\v s}} D.,
  {Springel} V., {Hernquist} L., 2013, \mnras, 429, 3353

\bibitem[{{Noeske} {et~al}\mbox{.}(2007){Noeske}, {Weiner}, {Faber},
  {Papovich}, {Koo}, {Somerville}, {Bundy}, {Conselice}, {Newman},
  {Schiminovich}, {Le Floc'h}, {Coil}, {Rieke}, {Lotz}, {Primack}, {Barmby},
  {Cooper}, {Davis}, {Ellis}, {Fazio}, {Guhathakurta}, {Huang}, {Kassin},
  {Martin}, {Phillips}, {Rich}, {Small}, {Willmer}, \&
  {Wilson}}]{2007ApJ...660L..43N}
{Noeske} K.~G. {et~al.}, 2007, \apjl, 660, L43

\bibitem[{{Oppenheimer} \& {Dav{\'e}}(2006)}]{2006MNRAS.373.1265O}
{Oppenheimer} B.~D., {Dav{\'e}} R., 2006, \mnras, 373, 1265

\bibitem[{{Pakmor} {et~al}\mbox{.}(2016){Pakmor}, {Springel}, {Bauer}, {Mocz},
  {Munoz}, {Ohlmann}, {Schaal}, \& {Zhu}}]{2016MNRAS.455.1134P}
{Pakmor} R., {Springel} V., {Bauer} A., {Mocz} P., {Munoz} D.~J., {Ohlmann}
  S.~T., {Schaal} K., {Zhu} C., 2016, \mnras, 455, 1134

\bibitem[{{Puchwein} \& {Springel}(2013)}]{2013MNRAS.428.2966P}
{Puchwein} E., {Springel} V., 2013, \mnras, 428, 2966

\bibitem[{{Rahmati} {et~al}\mbox{.}(2013){Rahmati}, {Pawlik}, {Rai{\v c}evic},
  \& {Schaye}}]{2013MNRAS.430.2427R}
{Rahmati} A., {Pawlik} A.~H., {Rai{\v c}evic} M., {Schaye} J., 2013, \mnras,
  430, 2427

\bibitem[{{Renaud} {et~al}\mbox{.}(2014){Renaud}, {Bournaud}, {Kraljic}, \&
  {Duc}}]{2014MNRAS.442L..33R}
{Renaud} F., {Bournaud} F., {Kraljic} K., {Duc} P.-A., 2014, \mnras, 442, L33

\bibitem[{{Rodighiero} {et~al}\mbox{.}(2011){Rodighiero}, {Daddi},
  {Baronchelli}, {Cimatti}, {Renzini}, {Aussel}, {Popesso}, {Lutz}, {Andreani},
  {Berta}, {Cava}, {Elbaz}, {Feltre}, {Fontana}, {F{\"o}rster Schreiber},
  {Franceschini}, {Genzel}, {Grazian}, {Gruppioni}, {Ilbert}, {Le Floch},
  {Magdis}, {Magliocchetti}, {Magnelli}, {Maiolino}, {McCracken}, {Nordon},
  {Poglitsch}, {Santini}, {Pozzi}, {Riguccini}, {Tacconi}, {Wuyts}, \&
  {Zamorani}}]{2011ApJ...739L..40R}
{Rodighiero} G. {et~al.}, 2011, \apjl, 739, L40

\bibitem[{{Rodriguez-Gomez} {et~al}\mbox{.}(2015){Rodriguez-Gomez}, {Genel},
  {Vogelsberger}, {Sijacki}, {Pillepich}, {Sales}, {Torrey}, {Snyder},
  {Nelson}, {Springel}, {Ma}, \& {Hernquist}}]{2015MNRAS.449...49R}
{Rodriguez-Gomez} V. {et~al.}, 2015, \mnras, 449, 49

\bibitem[{{Sales} {et~al}\mbox{.}(2015){Sales}, {Vogelsberger}, {Genel},
  {Torrey}, {Nelson}, {Rodriguez-Gomez}, {Wang}, {Pillepich}, {Sijacki},
  {Springel}, \& {Hernquist}}]{2015MNRAS.447L...6S}
{Sales} L.~V. {et~al.}, 2015, \mnras, 447, L6

\bibitem[{{Sanders} {et~al}\mbox{.}(1991){Sanders}, {Scoville}, \&
  {Soifer}}]{1991ApJ...370..158S}
{Sanders} D.~B., {Scoville} N.~Z., {Soifer} B.~T., 1991, \apj, 370, 158

\bibitem[{{Sargent} {et~al}\mbox{.}(2014){Sargent}, {Daddi}, {B{\'e}thermin},
  {Aussel}, {Magdis}, {Hwang}, {Juneau}, {Elbaz}, \& {da
  Cunha}}]{2014ApJ...793...19S}
{Sargent} M.~T. {et~al.}, 2014, \apj, 793, 19

\bibitem[{{Schaller} {et~al}\mbox{.}(2015){Schaller}, {Frenk}, {Bower},
  {Theuns}, {Jenkins}, {Schaye}, {Crain}, {Furlong}, {Dalla Vecchia}, \&
  {McCarthy}}]{2015MNRAS.451.1247S}
{Schaller} M. {et~al.}, 2015, \mnras, 451, 1247

\bibitem[{{Schaye} {et~al}\mbox{.}(2015){Schaye}, {Crain}, {Bower}, {Furlong},
  {Schaller}, {Theuns}, {Dalla Vecchia}, {Frenk}, {McCarthy}, {Helly},
  {Jenkins}, {Rosas-Guevara}, {White}, {Baes}, {Booth}, {Camps}, {Navarro},
  {Qu}, {Rahmati}, {Sawala}, {Thomas}, \& {Trayford}}]{2015MNRAS.446..521S}
{Schaye} J. {et~al.}, 2015, \mnras, 446, 521

\bibitem[{{Scoville} {et~al}\mbox{.}(2016){Scoville}, {Sheth}, {Aussel},
  {Vanden Bout}, {Capak}, {Bongiorno}, {Casey}, {Murchikova}, {Koda},
  {{\'A}lvarez-M{\'a}rquez}, {Lee}, {Laigle}, {McCracken}, {Ilbert}, {Pope},
  {Sanders}, {Chu}, {Toft}, {Ivison}, \& {Manohar}}]{2015arXiv151105149S}
{Scoville} N. {et~al.}, 2016, \apj, 820, 83

\bibitem[{{Scoville}(2013)}]{2013seg..book..491S}
{Scoville} N.~Z., 2013, {Evolution of star formation and gas}, Cambridge
  University Press, p. 491

\bibitem[{{Sijacki} {et~al}\mbox{.}(2007){Sijacki}, {Springel}, {Di Matteo}, \&
  {Hernquist}}]{2007MNRAS.380..877S}
{Sijacki} D., {Springel} V., {Di Matteo} T., {Hernquist} L., 2007, \mnras, 380,
  877

\bibitem[{{Snyder} {et~al}\mbox{.}(2015){Snyder}, {Torrey}, {Lotz}, {Genel},
  {McBride}, {Vogelsberger}, {Pillepich}, {Nelson}, {Sales}, {Sijacki},
  {Hernquist}, \& {Springel}}]{2015MNRAS.454.1886S}
{Snyder} G.~F. {et~al.}, 2015, \mnras, 454, 1886

\bibitem[{{Sparre} {et~al}\mbox{.}(2015{\natexlab{a}}){Sparre}, {Hayward},
  {Feldmann}, {Faucher-Gigu{\`e}re}, {Muratov}, {Kere{\v s}}, \&
  {Hopkins}}]{2015arXiv151003869S}
{Sparre} M., {Hayward} C.~C., {Feldmann} R., {Faucher-Gigu{\`e}re} C.-A.,
  {Muratov} A.~L., {Kere{\v s}} D., {Hopkins} P.~F., 2015{\natexlab{a}}, ArXiv:
  1510.03869

\bibitem[{{Sparre} {et~al}\mbox{.}(2015{\natexlab{b}}){Sparre}, {Hayward},
  {Springel}, {Vogelsberger}, {Genel}, {Torrey}, {Nelson}, {Sijacki}, \&
  {Hernquist}}]{2015MNRAS.447.3548S}
{Sparre} M. {et~al.}, 2015{\natexlab{b}}, \mnras, 447, 3548

\bibitem[{{Speagle} {et~al}\mbox{.}(2014){Speagle}, {Steinhardt}, {Capak}, \&
  {Silverman}}]{2014ApJS..214...15S}
{Speagle} J.~S., {Steinhardt} C.~L., {Capak} P.~L., {Silverman} J.~D., 2014,
  \apjs, 214, 15

\bibitem[{{Springel}(2010)}]{2010MNRAS.401..791S}
{Springel} V., 2010, \mnras, 401, 791

\bibitem[{{Springel}(2015)}]{2015ascl.soft02003S}
{Springel} V., 2015, {NGenIC: Cosmological structure initial conditions}.
  Astrophysics Source Code Library, ascl:1502.003

\bibitem[{{Springel} {et~al}\mbox{.}(2005){Springel}, {Di Matteo}, \&
  {Hernquist}}]{2005MNRAS.361..776S}
{Springel} V., {Di Matteo} T., {Hernquist} L., 2005, \mnras, 361, 776

\bibitem[{{Springel} \& {Hernquist}(2003)}]{2003MNRAS.339..289S}
{Springel} V., {Hernquist} L., 2003, \mnras, 339, 289

\bibitem[{{Stinson} {et~al}\mbox{.}(2013){Stinson}, {Brook}, {Macci{\`o}},
  {Wadsley}, {Quinn}, \& {Couchman}}]{2013MNRAS.428..129S}
{Stinson} G.~S., {Brook} C., {Macci{\`o}} A.~V., {Wadsley} J., {Quinn} T.~R.,
  {Couchman} H.~M.~P., 2013, \mnras, 428, 129

\bibitem[{{Suresh} {et~al}\mbox{.}(2015){Suresh}, {Bird}, {Vogelsberger},
  {Genel}, {Torrey}, {Sijacki}, {Springel}, \&
  {Hernquist}}]{2015MNRAS.448..895S}
{Suresh} J., {Bird} S., {Vogelsberger} M., {Genel} S., {Torrey} P., {Sijacki}
  D., {Springel} V., {Hernquist} L., 2015, \mnras, 448, 895

\bibitem[{{Teyssier}(2002)}]{2002A&A...385..337T}
{Teyssier} R., 2002, \aap, 385, 337

\bibitem[{{Teyssier} {et~al}\mbox{.}(2010){Teyssier}, {Chapon}, \&
  {Bournaud}}]{2010ApJ...720L.149T}
{Teyssier} R., {Chapon} D., {Bournaud} F., 2010, \apjl, 720, L149

\bibitem[{{Torrey} {et~al}\mbox{.}(2015){Torrey}, {Snyder}, {Vogelsberger},
  {Hayward}, {Genel}, {Sijacki}, {Springel}, {Hernquist}, {Nelson}, {Kriek},
  {Pillepich}, {Sales}, \& {McBride}}]{2015MNRAS.447.2753T}
{Torrey} P. {et~al.}, 2015, \mnras, 447, 2753

\bibitem[{{Torrey} {et~al}\mbox{.}(2012){Torrey}, {Vogelsberger}, {Sijacki},
  {Springel}, \& {Hernquist}}]{2012MNRAS.427.2224T}
{Torrey} P., {Vogelsberger} M., {Sijacki} D., {Springel} V., {Hernquist} L.,
  2012, \mnras, 427, 2224

\bibitem[{{Tremmel} {et~al}\mbox{.}(2015){Tremmel}, {Governato}, {Volonteri},
  \& {Quinn}}]{2015MNRAS.451.1868T}
{Tremmel} M., {Governato} F., {Volonteri} M., {Quinn} T.~R., 2015, \mnras, 451,
  1868

\bibitem[{{Vogelsberger} {et~al}\mbox{.}(2013){Vogelsberger}, {Genel},
  {Sijacki}, {Torrey}, {Springel}, \& {Hernquist}}]{2013MNRAS.436.3031V}
{Vogelsberger} M., {Genel} S., {Sijacki} D., {Torrey} P., {Springel} V.,
  {Hernquist} L., 2013, \mnras, 436, 3031

\bibitem[{{Vogelsberger} {et~al}\mbox{.}(2014{\natexlab{a}}){Vogelsberger},
  {Genel}, {Springel}, {Torrey}, {Sijacki}, {Xu}, {Snyder}, {Bird}, {Nelson},
  \& {Hernquist}}]{2014Natur.509..177V}
{Vogelsberger} M. {et~al.}, 2014{\natexlab{a}}, \nat, 509, 177

\bibitem[{{Vogelsberger} {et~al}\mbox{.}(2014{\natexlab{b}}){Vogelsberger},
  {Genel}, {Springel}, {Torrey}, {Sijacki}, {Xu}, {Snyder}, {Nelson}, \&
  {Hernquist}}]{2014MNRAS.444.1518V}
{Vogelsberger} M. {et~al.}, 2014{\natexlab{b}}, \mnras, 444, 1518

\bibitem[{{Weisz} {et~al}\mbox{.}(2012){Weisz}, {Johnson}, {Johnson},
  {Skillman}, {Lee}, {Kennicutt}, {Calzetti}, {van Zee}, {Bothwell},
  {Dalcanton}, {Dale}, \& {Williams}}]{2012ApJ...744...44W}
{Weisz} D.~R. {et~al.}, 2012, \apj, 744, 44

\bibitem[{{Wellons} {et~al}\mbox{.}(2015){Wellons}, {Torrey}, {Ma},
  {Rodriguez-Gomez}, {Vogelsberger}, {Kriek}, {van Dokkum}, {Nelson}, {Genel},
  {Pillepich}, {Springel}, {Sijacki}, {Snyder}, {Nelson}, {Sales}, \&
  {Hernquist}}]{2015MNRAS.449..361W}
{Wellons} S. {et~al.}, 2015, \mnras, 449, 361

\end{thebibliography}
}

\label{lastpage}

\end{document}